\documentclass[traditabstract]{aa} 
\usepackage{graphicx}
\usepackage{txfonts}
\usepackage{lscape}

\newcommand {\halpha} {\mbox{H$\alpha$} }
\newcommand {\sii} {[\ion{S}{II}] }

\begin{document}

\title{Star Formation in the Gulf of Mexico}

   \author{T. Armond
          \inst{1}
          \fnmsep\thanks{Part of this work was performed at Centro de Astrofisica da Universidade do Porto,
                         Rua das Estrelas, 4150-762 Porto, Portugal}
          \and
          B. Reipurth
          \inst{2}
          \and
          J. Bally
          \inst{3}
          \and
          C. Aspin
          \inst{2}
          \fnmsep\thanks{Visiting Astronomer at the Infrared Telescope Facility
which is operated by the University of Hawaii under contract to the
National Aeronautics and Space Administration.}
          }

   \institute{SOAR Telescope, Casilla 603, La Serena, Chile\\
              \email{tarmond@ctio.noao.edu}
         \and
             Institute for Astronomy, University of Hawaii at Manoa,
             640 N. Aohoku Place, Hilo, HI 96720, USA\\
              \email{reipurth@ifa.hawaii.edu, caa@ifa.hawaii.edu}
         \and
             Center for Astrophysics and Space Astronomy,
             University of Colorado, Boulder, CO 80309, USA\\
             \email{john.bally@colorado.edu}
             }

   \date{Received 09 June 2009; Accepted xx xxxx xxxx}

   \abstract{We present an optical/infrared study of the dense
     molecular cloud, L935, dubbed ``The Gulf of Mexico'', which
     separates the North America and the Pelican nebulae, and we
     demonstrate that this area is a very active star forming region.
     A wide-field imaging study with interference filters has revealed
     35 new Herbig-Haro objects in the Gulf of Mexico. A grism survey
     has identified 41 H$\alpha$ emission-line stars, 30 of them new.
     A small cluster of partly embedded pre-main sequence stars is
     located around the known LkH$\alpha$ 185-189 group of stars,
     which includes the recently erupting FUor HBC~722.}


\maketitle
%

\section{Introduction}

The \object{North America Nebula} (NGC 7000) and the adjacent
\object{Pelican Nebula} (IC 5070), both well known
for the characteristic shapes that have given rise to their names, are
part of the single large HII region W80 (Morgan et al. 1955;
Westerhout 1958). The central part of W80 is obscured by a large dust
cloud (L935), that defines the ``Atlantic Coast'' and the ``Gulf of
Mexico'' of the North America Nebula (Herbig 1958). Bally \& Scoville
(1980) modeled W80 as an expanding molecular shell, a cloud being
disrupted by early type stars born inside. For an overview of the
region, see the review by Reipurth \& Schneider (2008).

The distance to W80 has been the subject of some debate 
(Wendker 1968; Neckel et al. 1980; Armandroff \& Herbst 1981),
ranging from values of 500 pc to 1 kpc. We here adopt the commonly
accepted distance of 550 $\pm$ 50 pc as estimated by Laugalys 
et al. (2006). This distance is consistent with the estimates of 
Herbig (1958), Wendker et al. (1983), Strai\v{z}ys et al. (1993)
and Laugalys \& Strai\v{z}ys (2002).

Many authors have made searches for the ionizing sources of W80, but
until recently none were conclusive (e.g. Osterbrock 1957; Neckel et
al.  1980; Bally \& Scoville 1980). Comer\'on \& Pasquali (2005) have
finally found a good candidate among 2MASS detections in the cloud.
They proposed that the exciting source is an O5V star (2MASS
J205551.25+435224.6) located close to the geometric center of the
complex. Strai\v{z}ys \& Laugalys (2008) identified a few more
possibly highly reddened O-type stars that contribute to the
ionization of the North America and Pelican nebulae.
Figure~1 shows all of the region and the location of the 
exciting sources of the complex.

\begin{figure}
\centering
\resizebox{\hsize}{!}{\includegraphics{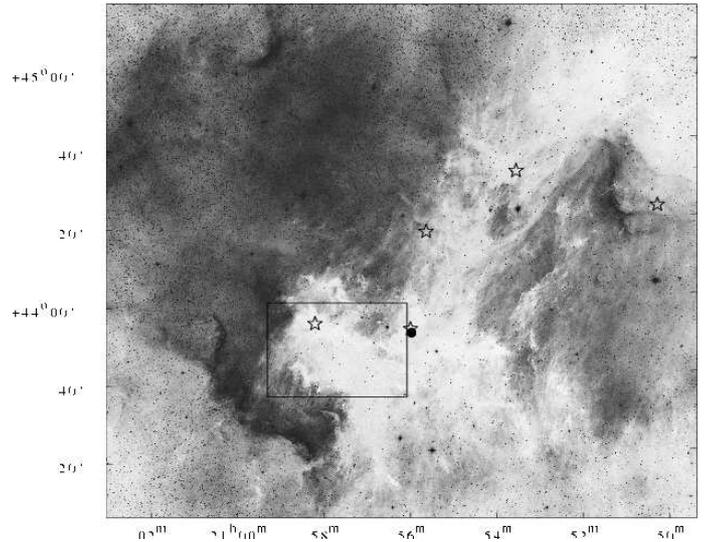}}
\caption{The North America and Pelican Nebula region, and the dark
cloud that divides them, L935, in a DSS image. 
The black circle marks the position of 
the exciting source proposed by Comer\'on \& Pasquali (2005).
The asterisks mark the position of other five candidate O-type stars 
from Strai\v{z}ys \& Laugalys (2008).
The rectangle in the Gulf of Mexico region shows the area 
surveyed at Subaru telescope, corresponding to the field shown in Fig. 2.
All coordinates are given in the equatorial J2000.0 system.}
\label{fig1}
\end{figure}

In a grism survey of the W80 region, Herbig (1958) detected
a population of H$\alpha$ emission-line stars, LkH$\alpha$~131 - 195,
mostly T Tauri stars, including the little group LkH$\alpha$ 185 to
189 located within the dark lane of the Gulf of Mexico, thus
demonstrating that low-mass star formation has recently taken place
here.

\begin{table}
\caption{Previously known \halpha emission-line stars in the Gulf of Mexico.}
\label{tab1}
\centering
\begin{tabular}{lllll}
\hline\hline
LK\halpha & GCVS & CoKu & HBC & KW \\ 
$^{(1)}$ & $^{(2)}$ & $^{(3)}$ & $^{(4)}$ & $^{(5)}$ \\
\hline
185 & V1539Cyg &                         & 720 & 53-9  \\
    &           & NGC7000/IC5070 IRS3    &     & 53-11 \\
    &           & NGC7000/IC5070 IRS4    &     & 53-13 \\
    &           & NGC7000/IC5070 IRS5    & 721 & 53-14 \\
    &           & NGC7000/IC5070 IRS6    &     & 53-19 \\
    &           & Lk\halpha 188 G5       &     & 53-17 \\
    &           & Lk\halpha 188 G4       & 722 & 53-18 \\
    &           & Lk\halpha 188 G3       &     & 53-20 \\
    &           & Lk\halpha 188 G2       &     & 53-22 \\
    &           & Lk\halpha 188 G1       &     & 53-23 \\
186 &           &                        & 723 & 53-24 \\
187 &           &                        & 724 & 53-25 \\
188 & V521Cyg  &                         & 299 & 53-26 \\
189 &           &                        & 725 & 53-27 \\

\hline
\end{tabular}
\begin{list}{}{}
\item[$^{\mathrm{*}}$] References: (1) Herbig 1958; (2) General Catalogue of Variable Stars; 
(3) Cohen \& Kuhi 1979; (4) Herbig \& Bell 1988; (5) Kohoutek \& Wehmeyer 1999.
\end{list}
\end{table}

Lk\halpha 188 was included in the Second Catalog of Emission-Line
Stars from Herbig \& Rao (1972) as HRC~299.
In a subsequent work, Welin (1973) detected only Lk\halpha 185 and 189
among \halpha emission-line stars in NGC 7000. 
Cohen \& Kuhi (1979) optically identified a group of five faint stars
associated with the small group containing Lk\halpha 186 to 189,
designating them as Lk\halpha 188~G1 to G5
(although they are actually closer to Lk\halpha 186). 
Infrared sources were also identified by Cohen \& Kuhi 
near the optical group, and were designated as NGC~7000/IC~5070~IRS~3
to IRS~6, with IRS~5 showing \halpha emission.
All of these \halpha emission-line stars were also included in
the catalogs of Herbig \& Bell (1988) and
Kohoutek \& Wehmeyer (1999). 
Table~1 lists the \halpha emission-line stars known in the
Gulf of Mexico prior to the present study.

Laugalys et al. (2006) made a photometric survey of the dark cloud 
L935, estimating spectral types, color indices and distances
for hundreds of stars. They used a photometric method to infer 
\halpha emission, listing 40 stars as possible \halpha emitters
in the total area surveyed ($\sim 1.2$ square degrees).
Later, the same group (Corbally et al. 2009) made a spectral 
analysis of the suspected young stellar objects (YSO) in
the North America/Pelican region, confirming the \halpha 
emission line in 19 stars.

In a recent study, Guieu et al. (2009) have used the {\it Spitzer Space
Telescope} with IRAC to identify more than 1600 YSO candidates in or
near the extended L935 cloud. 
They identify clusters and suggest that the region of the Gulf 
of Mexico contains the youngest stars of the complex.

  In August 2010 the star LkHa 188 G4 = HBC 722 has increased in
  brightness by more than 4 magnitudes and appears to be a new FU
  Orionis star, as reported by Semkov et al. (2010), Miller et al. 
  (2011), and Aspin et al. (2011).

A few surveys for Herbig-Haro (HH) objects have been made in the W80 region
(Ogura et al. 2002; Bally \& Reipurth 2003), resulting in the
identification of a number of outflows near the bright rim of the
Pelican Nebula.  But so far no search for HH objects has been
performed in the dark cloud of the Gulf of Mexico.

We present a survey for Herbig-Haro objects, for
H$\alpha$ emission-line stars, and for near-infrared sources in the
Gulf of Mexico, and demonstrate that this molecular cloud complex is very
rich in these signatures of current and recent low-mass star formation.

\begin{figure}
\centering
\resizebox{\hsize}{!}{\includegraphics{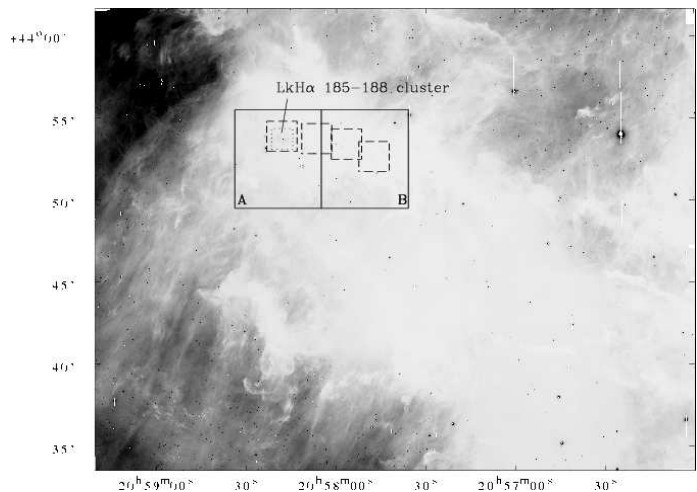}}
\caption{Portion of the L935 dark cloud corresponding to the 
Gulf of Mexico, on the \halpha image from the Subaru telescope taken in 2006. 
The two 7\arcmin$\times$7\arcmin\ fields (A and B) observed 
at the University of Hawaii 2.2m telescope are marked in 
solid lines. The full 7\arcmin$\times$14\arcmin\ field 
corresponds to the area shown in Fig. 4. The four fields observed 
through an H$_2$ filter at UKIRT are marked with the dashed line 
squares. The dotted line square marks the $JHKL$ field observed
at IRTF, centered on the Lk\halpha 185-188 cluster.}
\label{fig2}
\end{figure}

\section{Observations}

We have used a set of optical and near-infrared observations for 
this survey, which is summarized in Table 2.

The University of Hawaii 2.2m telescope on Mauna Kea was used to
obtain a set of optical images of the region containing the group of
stars Lk\halpha 185 to 189.  Both broad (VRI) and narrow band (\sii
6717/6731\AA~and \halpha 6563\AA) filters were used.  The Wide Field
Grism Spectrograph was also used to obtain grism images with the
\halpha filter to detect \halpha emission in the stars through
slit-less spectroscopy.  Previous observations obtained by George
Herbig in 1998, and kindly put at our disposal, provided the
opportunity to compare the emission line equivalent widths at two
different epochs.

Near-infrared $JHKL$ images were obtained towards the center of 
the optical cluster using the NASA Infrared Telescope Facility 
(IRTF) at Mauna Kea. The instrument used was the NSFCAM, 
a 1 to 5~$\mu$m imager with a 256$\times$256 InSb detector.
Reduction was carried out with standard IRAF procedures 
and aperture photometry was obtained with APPHOT package.
The typical uncertainties for the $VRI$ and $JHKL$ magnitudes 
were of the order of 0.05 mag.

Near-infrared images were also obtained with the 4m United Kingdom 
Infrared Telescope (UKIRT) at Mauna Kea, using 
UKIRT Fast-Track Imager (UFTI), a 1 to 2.5~$\mu$m camera with a 
1024$\times$1024 HgCdTe array.
Narrow-band H$_2$ and [FeII] filters were used to observe selected regions. 
$JHK$ non-photometric images were obtained to help identify embedded sources. 
The images were reduced with the standard UKIRT ORAC-DR pipeline.

Deep wide-field images were taken with 
SuprimeCam on the 8m Subaru telescope, also using \halpha and \sii filters, 
with seeing in individual sub-exposures ranging from 0.50$''$ to 0.54$''$
and from 0.47$''$ to 0.56$''$, respectively. 
The \halpha image is shown in Fig. 2, where all the regions
surveyed are marked.

\begin{table*}
\caption{Observations.}
\label{tab2}
\centering
\begin{tabular}{llllll}
\hline\hline
Telescope/ & Date & Filters & Exp. Time & FOV$^*$   & Plate scale \\
Instrument &      &         &  (s)      & (\arcmin) & (\arcsec pix$^{-1}$) \\
\hline
UH 2.2m/        & 2002 Jun. 16-17 & V     & 1200 & 7$\times$14 (AB) & 0.22 \\
Tek CCD         &                 & R     &  900 & 7$\times$14 (AB) &      \\
                &                 & I     &  900 & 7$\times$7 (A)   &      \\
                &                 & [SII] & 5400 & 7$\times$7 (A)   &   \\
\hline
UH 2.2m/     & 2002 Jul. 14 & [SII]    & 5400 & 7$\times$7 (B)   & 0.34 \\
Tek CCD+WFGS &              & H$\alpha$& 2700 & 7$\times$14 (AB) &   \\
             &              & Grism    & 2700 & 7$\times$14 (AB) &   \\
             & 1998 Oct. 12 & Grism    &  900 & 7$\times$7 (A)   &     \\
\hline
UKIRT/ & 2002 Jul. 13 & J, H, K       & 300 & 3$\times$9                & 0.091 \\
UFTI   &              & H$_2$, [FeII] & 500 & 4 $\times$ 1.5$\times$1.5 &  \\
\hline
IRTF/  & 1999 Aug. 30 & J    & 90 &  1.25$\times$1.25 & 0.30 \\
NSFCAM &              & H, K & 30 &                   &      \\
       &              & L    & 40 &                   &      \\
\hline
SUBARU/     & 2006 May 27-28 & H$\alpha$ & 2700 & 34$\times$27 & 0.20 \\
Suprime Cam &                & [SII]     & 3000 &              &      \\

\hline
\end{tabular}
\begin{list}{}{}
\item[$^*$] -- The fields observed are marked in Fig. 2.
\end{list}
\end{table*}

\begin{figure*}
\includegraphics[width=18cm]{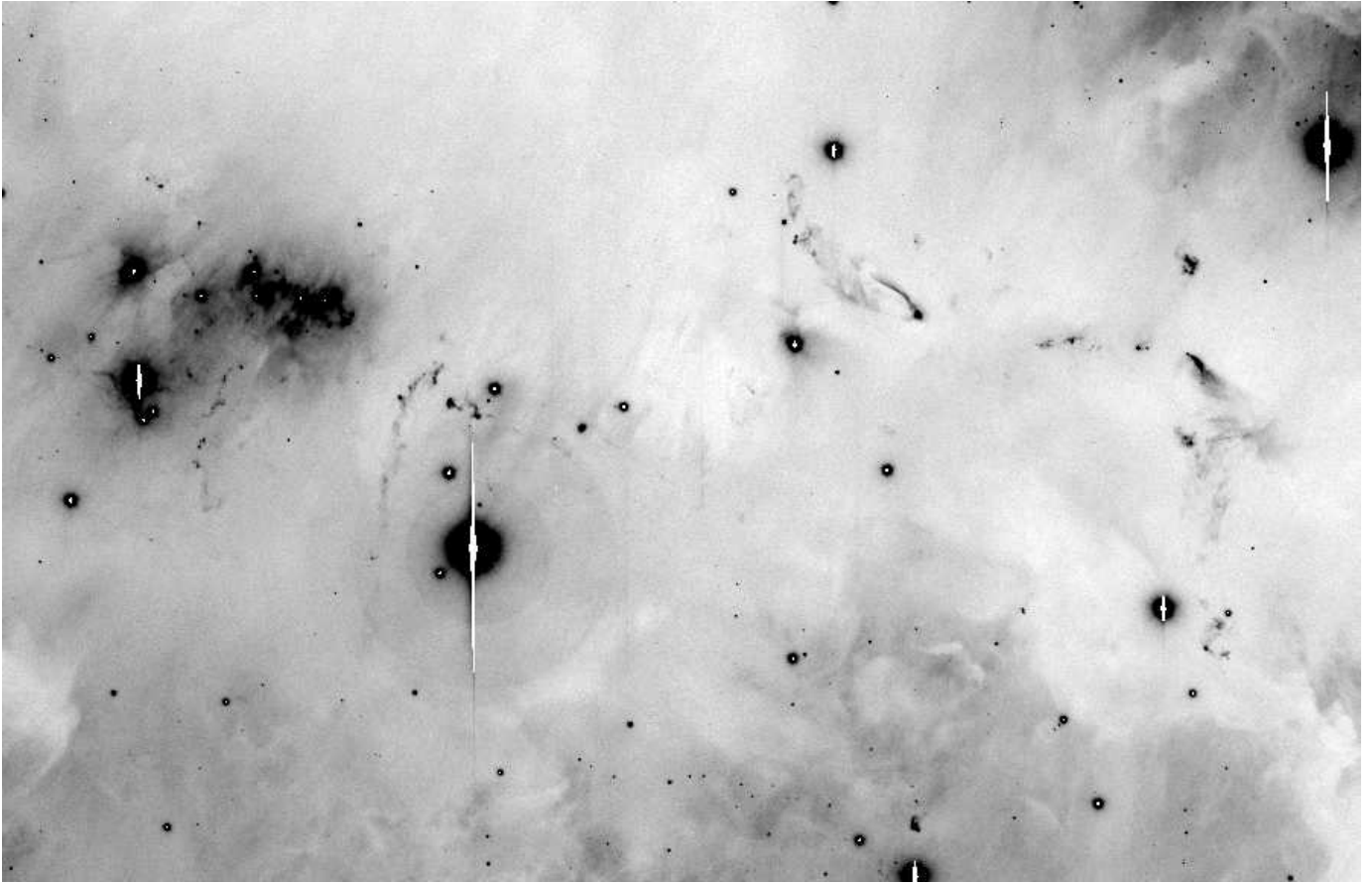}
\caption{A panorama of the most active region in the Gulf of Mexico
based on a \sii image from the Subaru telescope, showing most of the
Herbig-Haro objects detected. The image size is approximately
6\arcmin$\times$10\arcmin. North is up and East is left.
}
\label{fig3}
\end{figure*}

In images from the {\it Spitzer Space Telescope}
(Program ID $\#$20015: IRAC and MIPS observations of the North
America and Pelican Nebulae, PI: Luisa Rebull), we can see the 
region in the mid-infrared, with many embedded sources,
still invisible at near-infrared wavelengths.
Those archival images were used to derive IRAC 3.6, 4.5, 5.8 and
8~$\mu$m and MIPS 24~$\mu$m magnitudes for selected stars, 
given in Table 6.  

\section{New Herbig-Haro Flows}

\begin{figure*}
\centering
\resizebox{\hsize}{!}{\includegraphics{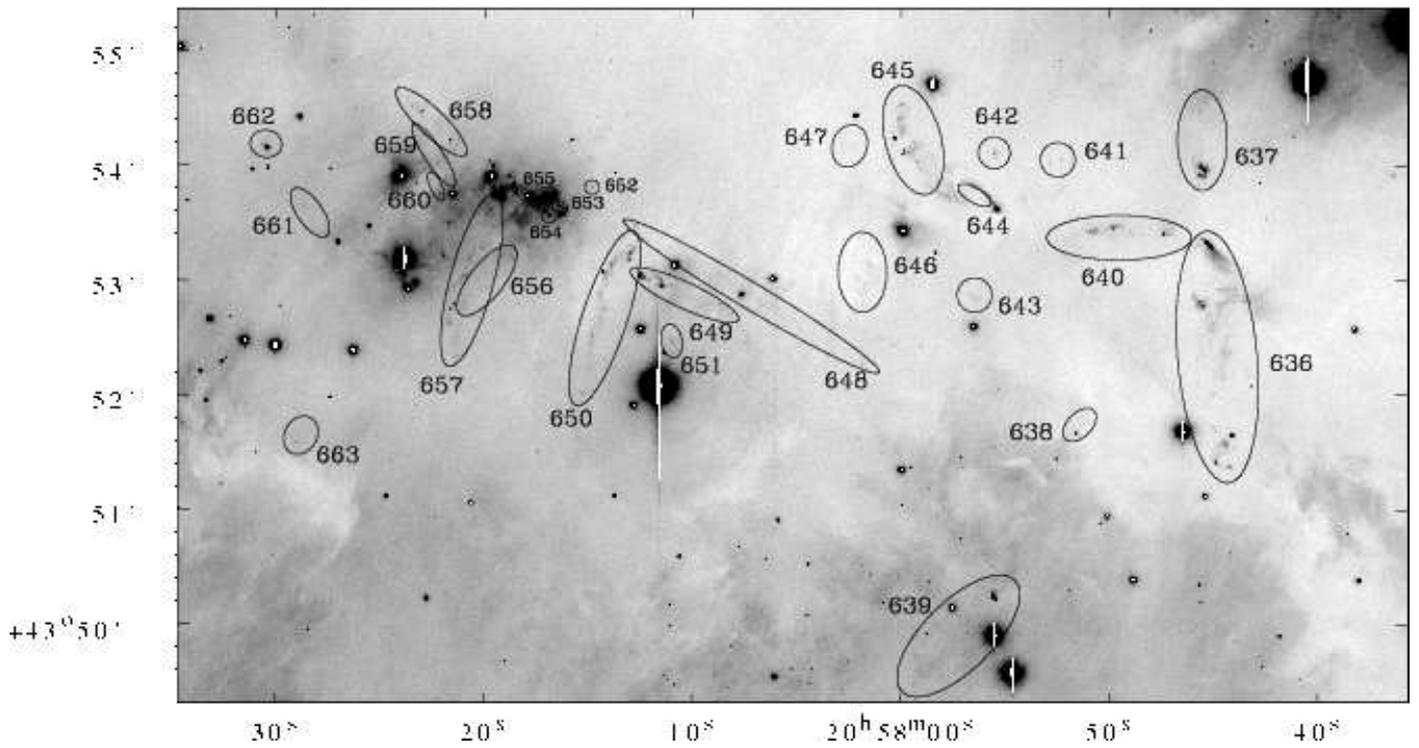}}
\caption{Identification of new HH objects in the LkH$\alpha$ 185-188
  cluster region. This is a \sii image obtained at the Subaru
  telescope. The figure shows the area of our original survey in
  2002.}
\label{fig4}
\end{figure*}

\begin{figure*}
\centering
\resizebox{\hsize}{!}{\includegraphics{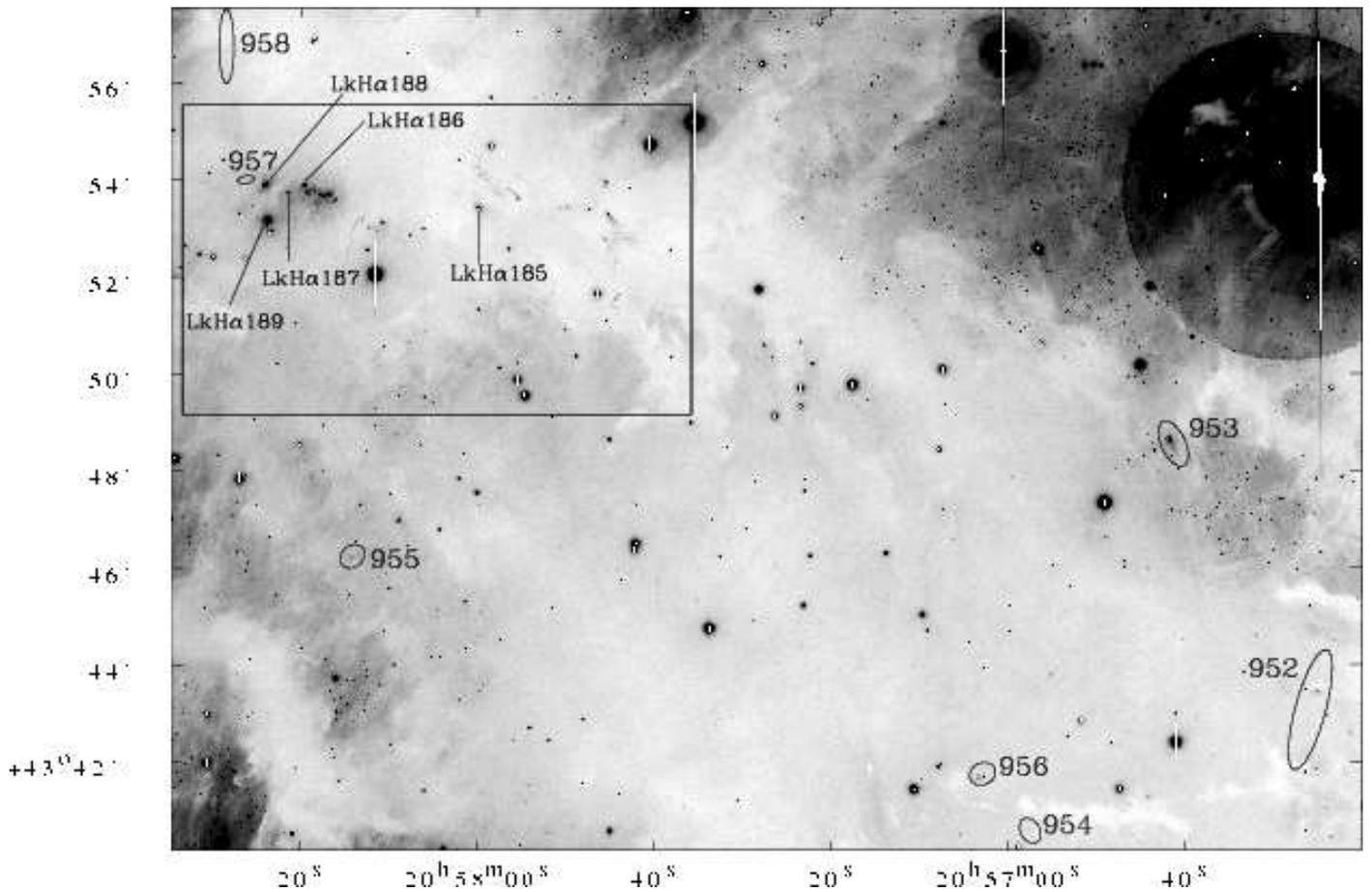}}
\caption{\sii image from the Subaru telescope showing the entire region surveyed for HH
objects. The rectangle shows the region seen in Fig. 4. The stars
LkH$\alpha$~185-189 are marked.}
\label{fig5}
\end{figure*}

\begin{figure*}
\centering
\includegraphics[width=17cm]{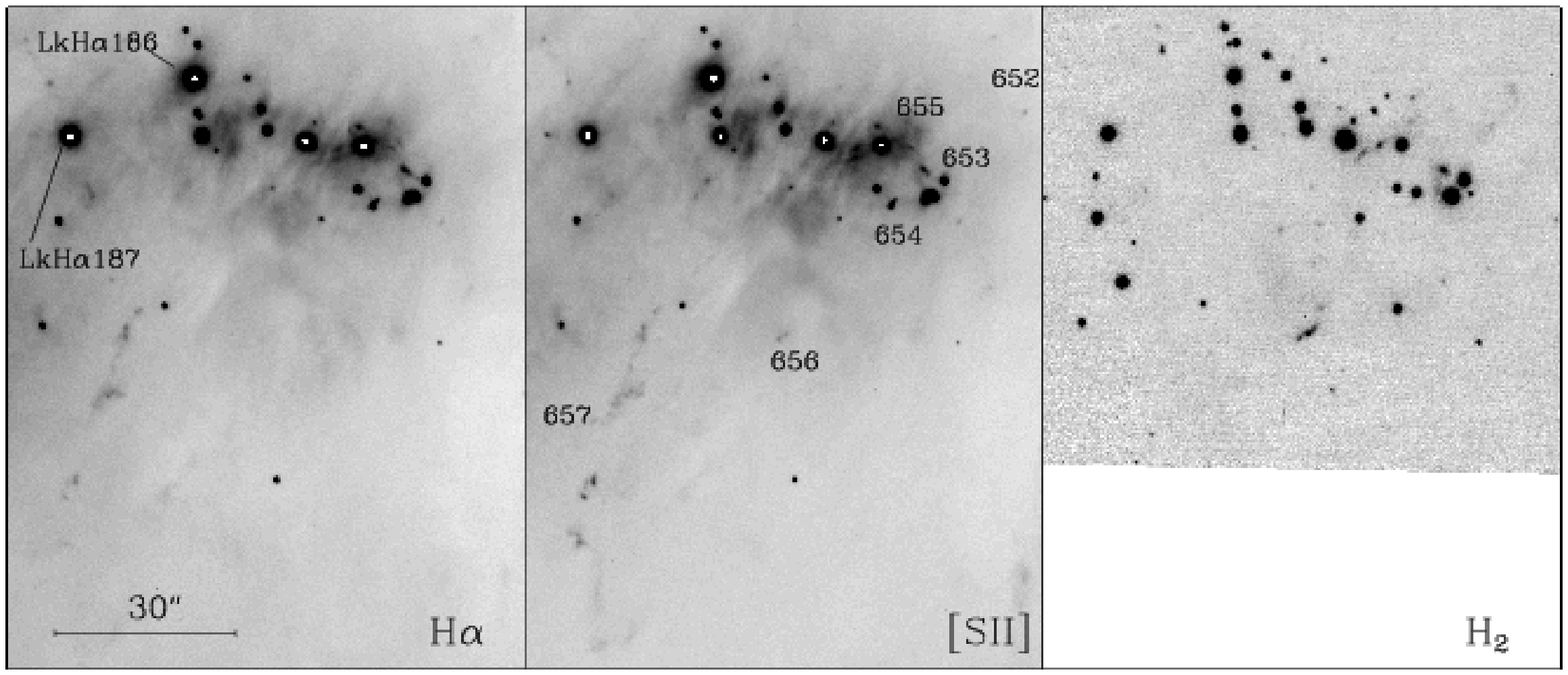}
\caption{The region of the optical cluster around Lk\halpha 186
and the flows in H$\alpha$, \sii and H$_2$ filters. 
We note that most of the flows are stronger in [\ion{S}{II}], but there 
are some also strong in H$_2$, like HH 655 and 656, which are more embedded.
To the northeast, an embedded flow in H$_2$ could be
related to the optical knots of HH 652.
In the H$_2$ image we also see the probable source of HH 654 just north of it,
as well as other embedded point sources.}
\label{fig6}
\end{figure*}

In the \sii and \halpha images obtained with the UH 2.2m telescope in
2002 we have found 28 HH objects.  Most of them appear only in the
\sii images; the few that are detected in \halpha are stronger in
[\ion{S}{II}], showing that these are low excitation shocked jet
material, and not photo-ionized nebulae.  Some of the flows were also
detected in the near-infrared H$_2$ images.  The broadband $I$ images
were also checked to prevent reflection nebulae to be identified as an
HH object.  The association of HH objects and H$_2$ knots with
specific stellar sources was done by visual inspection based on proximity and
apparent alignment.

\begin{table*}
\caption{Herbig-Haro objects in the Gulf of Mexico.}
\label{tab3}
\centering
\begin{tabular}{ccccccl}
\hline\hline
HH$^a$ & $\alpha$(J2000)$^b$ & $\delta$(J2000)$^b$ & [\ion{S}{II}]$^c$ & $\halpha ^c$ & H$_2^c$ & Comments$^d$ \\
\hline
636 &  20 57 45.5 & 43 52 47 & y & y & n & 2\arcmin ~long \\
637 &  20 57 45.5 & 43 53 58 & y & y & y & strong center, fainter structure 30\arcsec ~N\\
638 &  20 57 51.6 & 43 51 40 & y & y & - & strong knot plus filament \\
639 &  20 57 55.5 & 43 50 15 & y & y & - & strong bow-shock plus larger fainter structure 1\arcmin ~SE \\
640 &  20 57 47.3 & 43 53 25 & y & y & y & extends to E and W, 1\arcmin ~wide \\
641 &  20 57 52.7 & 43 54 02 & y & w & n & bow plus knot 7\arcsec ~NW \\
642 &  20 57 55.5 & 43 54 07 & y & y & n & knot plus stucture $\sim$7\arcsec ~wide \\
643 &  20 57 56.5 & 43 52 52 & y & w & y & $\sim$7\arcsec ~wide \\
644 &  20 57 55.4 & 43 53 37 & y & y & w & filament, extends 30\arcsec ~to NE \\
645 &  20 57 59.6 & 43 54 07 & y & y & n & extends to N and S, 40\arcsec ~wide \\
646 &  20 58 01.4 & 43 53 07 & y & y & n & knot plus bow, $\sim$25\arcsec ~wide \\
647 &  20 58 02.3 & 43 54 09 & w & n & n & weak, $\sim$20\arcsec ~wide \\
648 &  20 58 01.8 & 43 52 17 & y & w & n & 3\arcmin ~long chain of knots and filaments around HBC 721\\
649 &  20 58 12.5 & 43 53 03 & y & y & n & strong knot plus filaments 1\arcmin ~to SW \\
650 &  20 58 13.0 & 43 53 14 & y & y & n & curved, extends 1\farcm5 to S \\
651 &  20 58 11.0 & 43 52 28 & w & w & - & weak, 10\arcsec ~long \\
652 &  20 58 14.9 & 43 53 49 & w & n & w & weak knot \\
653 &  20 58 16.4 & 43 53 40 & y & y & n & strong, points away from MKH$\alpha$ 10 \\
654 &  20 58 16.9 & 43 53 33 & y & y & n & strong, stellar-like \\
655 &  20 58 17.1 & 43 53 47 & y & w & y & around star G4, 18\arcsec ~wide \\
656 &  20 58 18.6 & 43 53 11 & y & w & y & weak, extends 35\arcsec ~to SE \\
657 &  20 58 21.5 & 43 52 47 & y & w & n & 1\farcm3 long \\
658 &  20 58 21.7 & 43 54 14 & y & y & n & chain of knots 36\arcsec ~to NE \\
659 &  20 58 21.9 & 43 53 54 & y & y & n & knots, extends 30\arcsec ~to NE \\
660 &  20 58 22.2 & 43 53 45 & y & w & n & knot plus faint stucture 10\arcsec ~to NE \\
661 &  20 58 27.9 & 43 53 25 & y & w & - & knot plus larger sturcture 25\arcsec ~to NE \\
662 &  20 58 30.3 & 43 54 09 & y & n & - & two knots, W and NE of star MKH$\alpha$ 29 \\
663 &  20 58 29.0 & 43 51 35 & w & n & - & faint \\
\hline
952 &  20 56 23.2 & 43 43 55 & w & n & - & three weak knots across nebulous star \\
953 &  20 56 41.6 & 43 48 39 & y & y & - & very strong 24\arcsec ~wide \\
954 &  20 56 57.6 & 43 40 38 & w & n & - & 1\farcm2 SW from nebulous star \\
955 &  20 58 14.3 & 43 46 13 & y & n & - & triangular shape, 1\farcm5 SE from nebulous star \\
956 &  20 57 02.8 & 43 41 46 & y & n & - & knot near nebulous star \\
957 &  20 58 26.5 & 43 54 00 & w & n & - & weak jet-like 25\arcsec ~long, W from star MKH$\alpha$ 28 \\
958 &  20 58 28.3 & 43 56 44 & w & n & - & chain of four knots 1\arcmin ~long \\
\hline
p   &  20 56 21.0 & 43 47 33 & w & w & - & weak knots \\
p   &  20 57 21.0 & 43 49 28 & w & w & - & weak knots \\
p   &  20 58 06.4 & 43 53 01 & w & w & n & knot near MKH$\alpha$ 6 \\
p   &  20 58 08.0 & 43 40 27 & w & ? & - & 10\arcsec ~long structure, confusion with reflection area \\
p   &  20 58 24.1 & 43 50 43 & w & n & - & weak, similar to HH 663 \\

\hline
\end{tabular}
\begin{list}{}{}
\item[$^a$] -- HH 636 to 663 from 2002 observations, HH 952 to 958 from 2006 
observations, p means possible HH objects that need confirmation.
\item[$^b$] -- The positions are measured at the brightest point.
\item[$^c$] -- Indicates if the flow is observed at each filter. y means
yes, n means no, w means weak and - means it is out of the 
observed field in that filter.
\item[$^d$] -- Approximate description of how the HH object appears in the 
\sii image.
\end{list}
\end{table*}

In the \sii and \halpha images taken with the Subaru telescope in 2006, 
the amount of detail is greater and the field-of-view is much
larger than in the previous images. 
We have detected 7 additional new HH objects on these images. 
One of them lies in the region surveyed in 2002, but it was 
too faint to be identified then.

Figure 3 shows a high contrast \sii image from the Subaru telescope
with the region containing most of the Herbig-Haro objects found.  The
35 new HH objects and their identification are shown in Figs. 4 and 5,
which are also based on \sii images. In Figs. 6 to 11 we see each set
of flows in every narrow-band filter observed.  Additionally, stamps
of the new objects discovered in the wide field Subaru image are shown in Fig.
12, except HH 957, which is seen in Fig. 7.  Table~3 lists the
identifications, positions and a brief description of all the objects
detected.

\subsection{The region around the optical cluster of Lk\halpha 186}

In the region around the optical cluster the density of stars and
shocks is so high that identification of the driving sources of flows
is limited by confusion.  Additionally, the stars of the cluster are
surrounded by reflection nebulosity, especially the four brightest
stars Lk\halpha 186 to 189.

HH 652 (Fig. 6) is a single knot only seen in [\ion{S}{II}]. 
In the H$_2$ image, there is a flow nearby but we
cannot be sure if it is related to HH 652.  
 
HH 653 seems to point away from the star MK\halpha 10 (MK\halpha is the
designation for the new \halpha emission-line stars; see next section). 
HH~653 is strong in all the three narrow band images and also shows a 
weak continuum component.  
 
HH 654 is only detected in the optical narrow band images. It is very bright
and points away from a source only seen in H$_2$ and $JHK$.  
 
Star G4 (HBC~722)is surrounded by what we call HH 655, which consists of an
eastern flow, stronger in [\ion{S}{II}], but also very strong in
H$_2$, plus a flow to the west and a knot to the north of the star.  In the
H$_2$ image there is also a diffuse nebulosity to the north, but west of the
optical knot. With the present observations it is impossible to
determine if all are in fact related to star G4.
 
HH 656 is stronger in \sii than in H$\alpha$, and very strong in H$_2$.  
It lies at a position where it might be driven by a star faint in the 
optical, but bright in the near-infrared H$_2$ image, located 
20\arcsec\ to northwest.  
 
HH 657 is a long flow with curved appearance which seems to point back at G1.
But it is not entirely clear if the northern knots are really
connected to the flows in the center. And also it is not clear if the
southern part of the flow, resembling a bow shock, is related to HH
657 or to HH 656 or to none of them.

\subsection{The region around Lk\halpha 188-189}

Figures 4 and 7 show the region east of the optical cluster, where
Lk\halpha 188 and Lk\halpha 189 are very bright and immersed in
diffuse emission.  
 
HH 658 is a small chain of knots, strong in \sii and
H$\alpha$, pointing away from the direction of Lk\halpha 186. 

HH 659 and 660 are knots too, and both appear to come from 
Lk\halpha 187, in different directions.  
 
HH 661 has also two knots plus a fainter flow 25\arcsec\ away. There
is a possible counter-jet in \halpha just southwest from the possible
source, MK\halpha 24. Note that the fainter MK\halpha 26 also lies in
a position where it could be the source of HH 661, or at least the
nearest knots. 

HH 662 consists of two single knots around MK\halpha
29, not perfectly aligned through the star.  
 
The HH 957 weak jet was only distinguished with confidence in the 
recent Subaru images. It seems to point out of the faint star MK\halpha 27.

\begin{figure}
\centering
\includegraphics[width=7cm]{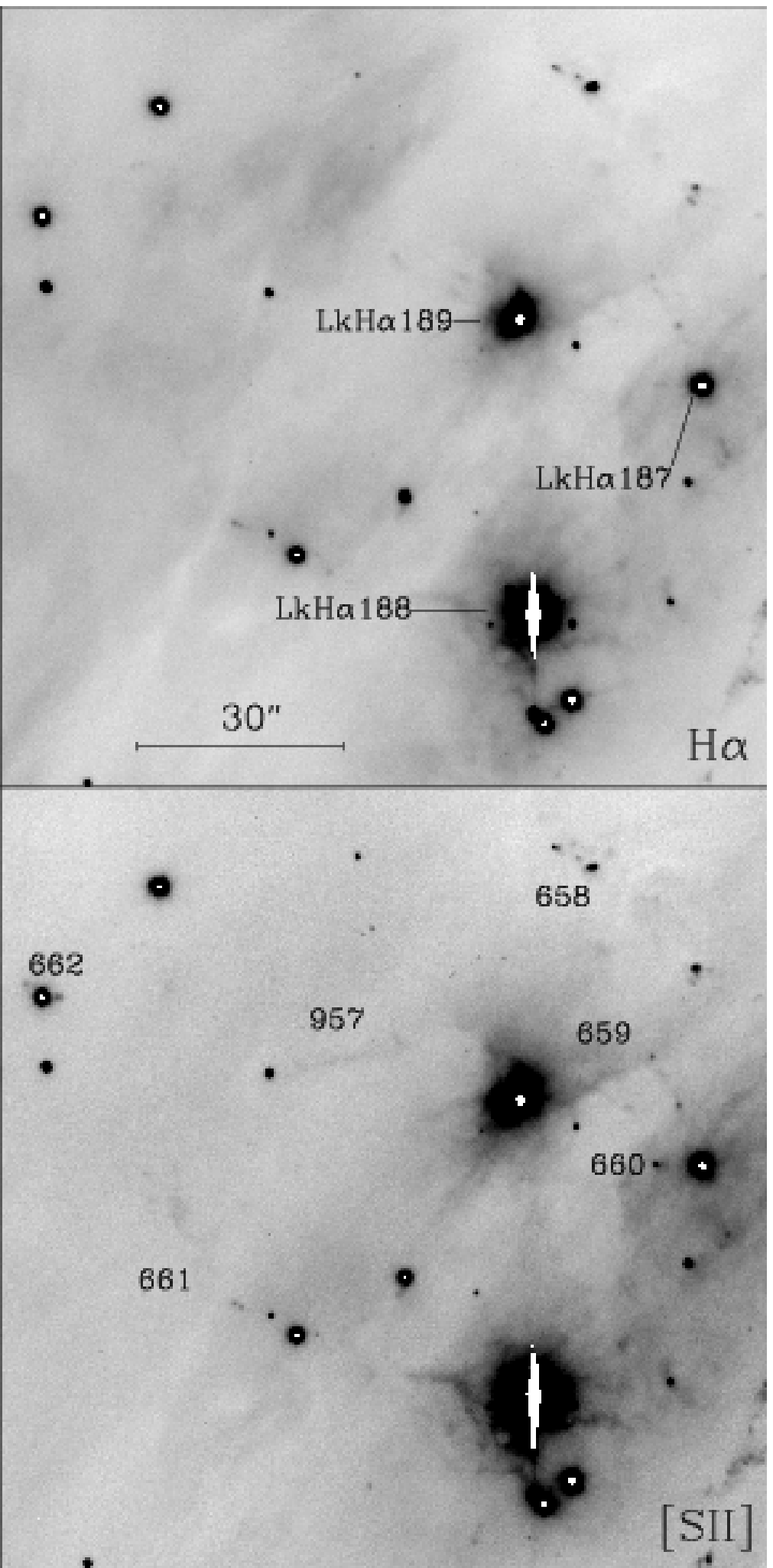}
\caption{East of the optical young cluster, Lk\halpha 188 and 189 are
surrounded by reflection nebulae. The HH objects found in the area are indicated
in the \sii image.
HH 658 to 662 are mainly knots aligned pointing away from a star.
The HH 957 weak jet was only distinguished with confidence in the recent Subaru images.}
\label{fig7}
\end{figure}

\begin{figure}
\centering
\includegraphics[width=8cm]{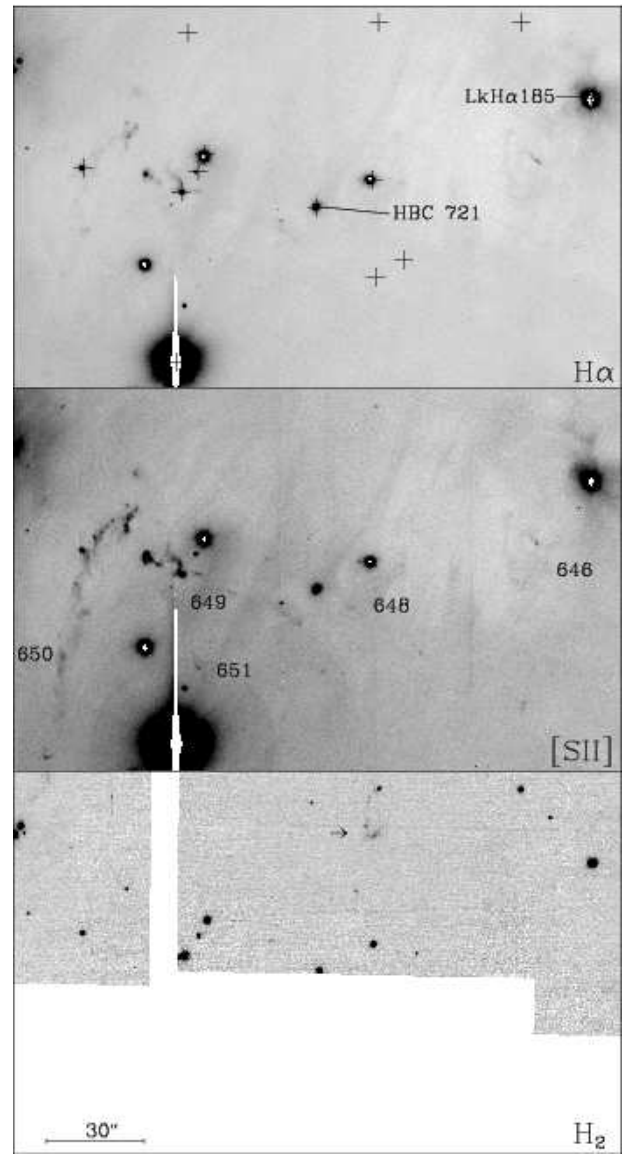}
\caption{Between the optical cluster and Lk\halpha 185,
the crosses in the \halpha image mark the positions of the
brightest sources in the 8~$\mu$m {\it Spitzer} images.
In the \sii image are marked the HH objects found.
In the H$_2$ image an embedded flow is seen, only in the near infrared.
Its probable source is barely seen in the optical images,
but it is bright in the infrared images.}
\label{fig8}
\end{figure}

\subsection{The region around HBC 721}

West of the optical cluster (Figs. 4 and 8) the star HBC 721
is in the center of a chain of knots and filaments called HH 648, 
which seems to be a bipolar jet. 
 
HH 646 is a faint structure and Lk\halpha 185 could be its source. 
 
HH 649 has a very strong knot, with a faint continuum component, 
a fainter filament pointing southwest, and another filamentary 
structure west of MK\halpha 8. It could be a bipolar jet from 
MK\halpha 8, although the filament is not perfectly aligned with it.  
Our near-infrared $JHK$ images show a fainter redder star east of 
MK\halpha 8 and the filament next to HH 649's head points directly to it. 
We also see two embedded sources to the west in the {\it Spitzer} images, 
which lie in the direction of this flow. The positions of the bright
{\it Spitzer} sources are marked with a cross in the figures.
  
HH 650 resembles HH 657 in structure, shape and orientation. It is a curved
flow, more than 1\farcm5 long.  A faint near-infrared
source is located at the northern tip of the flow. There is also a
bright infrared source (corresponding to IRS 6 from Cohen \& Kuhi 1979) $\sim
25$\arcsec\ south of the flow.

HH 651 is a faint knot seen both in \sii and \halpha and a fainter 
filament only seen in [\ion{S}{II}]. There is no indication of its source.

\begin{figure}
\centering
\includegraphics[width=8cm]{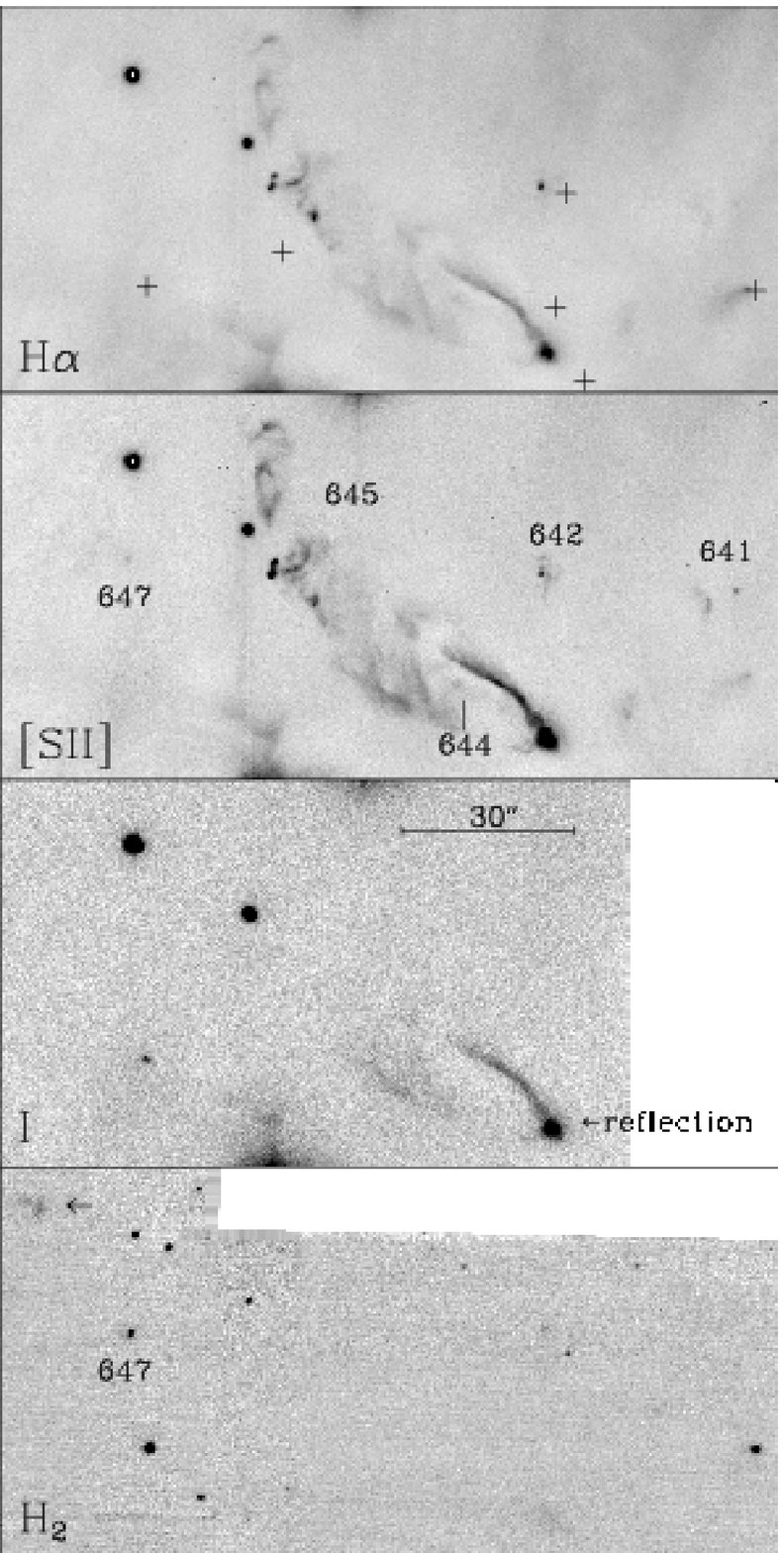}
\caption{The region around HH 644 and 645 in H$\alpha$, [\ion{S}{II}], $I$ and H$_2$.
The bright point source and the curved filaments are actually reflection
nebulae, visible at broadband $I$, and trace probably a cavity drawn by HH 644,
which is only seen in \sii and H$\alpha$.
HH 647 is much brighter in the H$_2$ image, where there is also a
nebulosity in the upper left corner of the image,
not associated with optical structures.
The strong {\it Spitzer} sources have their positions marked by a cross
in the \halpha image.}
\label{fig9}
\end{figure}

\subsection{A reflection cavity}

To the west (Fig. 9), the curved bright filament around HH 644 is due
to reflected light, probably the wall of a cavity formed by a
molecular outflow and illuminated by the embedded driving source.
Faint reflected light can also be seen from the opposing wall.
HH~644 is a knot plus a filament seen in H$\alpha$, stronger in
[\ion{S}{II}].  No probable source is seen in our optical or
near-infrared images. In the {\it Spitzer} images we can see many
embedded sources in the area, one coincides with the expected position
of a source for HH 644 just southwest of the cavity.
 
HH 645 is possibly an extension of HH 644, but we cannot be sure of
this without proper motion measurements. There are two {\it Spitzer}
sources, one just northeast of the probable source of HH 644, and
another source about 30\arcsec\ northeast which also could be related
to HH 645.  There are two faint emission patches stronger in \halpha
west of the reflection nebula which are probably HH objects associated
with a near infrared source seen in the H$_2$ image and also by {\it
  Spitzer}. We need confirmation of their nature, however, because
they are stronger in the \halpha images and they lie in a region where
we lack corresponding broadband $I$ images, so they could in principle
be parts of an illuminated cavity, and are therefore not given HH
numbers here.
 
HH~641 and HH~642 are diffuse structures, and there is one {\it Spitzer} 
source almost coinciding with HH 642. 
 
HH 647 is very faint in the optical, but strong in the H$_2$ image.
To its south there is one point source, faint in the optical, stronger
in near-infrared and even brighter in {\it Spitzer} images, and it
could be associated with this HH object. There are other probable
flows in H$_2$ in this region, as seen in Figs. 8 and 9.

\subsection{The westernmost flows}

The western flows HH 636, 637, 638 and 640 are shown in Fig. 10.
They are very bright in [\ion{S}{II}] and not so faint in H$\alpha$.
There is also an extended reflection nebula, bright in the broad band
$R$ images, probably an illuminated cavity. 
Unfortunately our $I$-band image does not cover this region.

Based on the geometry of the flows and the near-infrared appearance of
the point sources, HH 636 and 637 appear to be part of a bipolar flow
driven by the western star, visible in the H$_2$ image and also bright
in {\it Spitzer} images, marked A in Fig. 10.

The star to the east (B), also bright in {\it Spitzer} images and
showing signs of a cavity, is probably the source of HH 640, which
shows a small flow to the west and a larger flow to the east of the
star.  The H$_2$ images show components of HH 640 and HH 637.
 
HH 638 is a bright knot and a faint filament that point back to a 
red source seen in our near-infrared images and also in the 
{\it Spitzer} images, marked C in the figure.

\begin{figure}
\centering
\includegraphics[width=8.8cm]{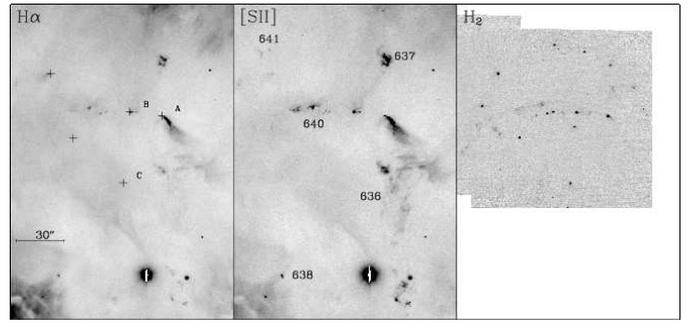}
\caption{The western-most flows in H$\alpha$, \sii and H$_2$.
In the H$_2$ image embedded flows and sources are seen.
The strong {\it Spitzer} sources have their positions marked by a cross.
Star marked A is a possible source for HH 636 and HH 637, B
is likely the source of HH 640 and C is possibly related to HH 638.}
\label{fig10}
\end{figure}

\begin{figure}
\centering
\includegraphics[width=8.5cm]{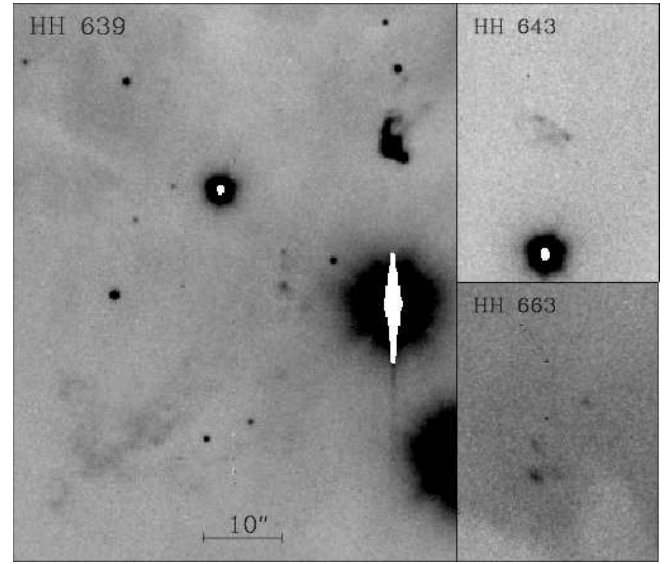}
\caption{HH 639, HH 643 and HH 663 in \sii images
at same scale.}
\label{fig11}
\end{figure}

In Fig. 11, HH 639 is a very bright bow and a fainter larger bow to
the southeast.  HH 643 is faint even in [\ion{S}{II}], visible in
H$_2$.  HH 663 is faint and visible only in the \sii image.  There are
no conclusive candidate sources for these HH objects.

\subsection{The newest flows}

Figure~12 shows 6 of the 7 HH objects identified in the new
wide-field Subaru images, most of them in the southwestern region of
the Gulf of Mexico, far from the region where the major activity was
detected first. 

HH 952 is a chain of three faint knots whose axis passes
through a nebulous star,  with an illuminated cavity.  

HH 953 is a bright object, with no candidate source identified.

HH 954 and HH 956 are faint objects found near a nebulous star.  

HH 955 is faint and located near a reflection nebula that resembles a
nearly edge-on disk. There is a source associated with this reflection
in the optical and infrared, but its 2MASS colors show no
infrared excess.

HH 958 is located north of the optical cluster, it is probably
related to nearby infrared sources seen by {\it Spitzer}.

\begin{figure*}
\centering
\includegraphics[width=18cm]{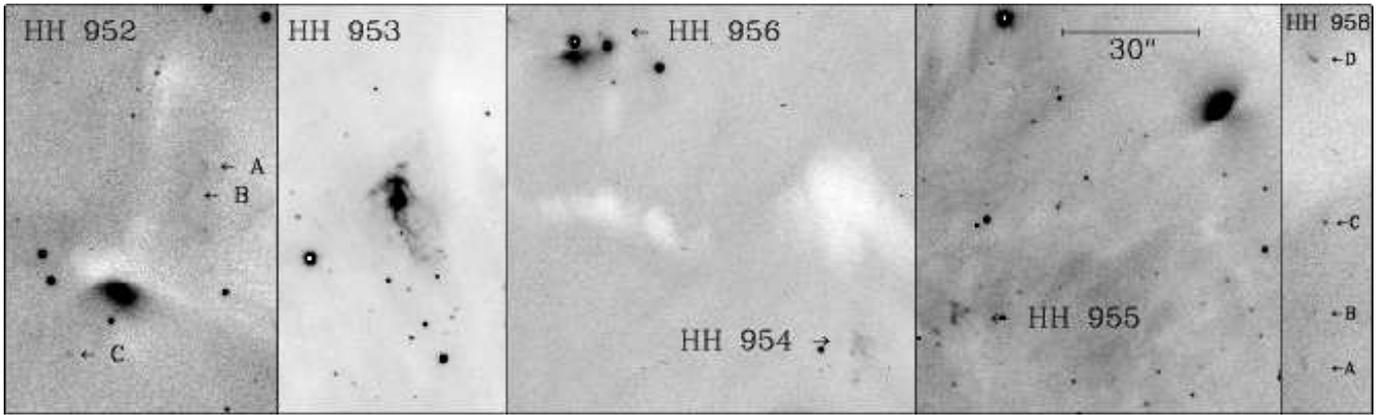}
\caption{The new HH objects found in the large-field Subaru \sii images.
All the images are at the same scale. HH 957 is displayed in Fig. 7.}
\label{fig12}
\end{figure*}

Other possible HH objects were found and are listed as ``p'' in 
Table~3, but further observations are needed to confirm their true nature,
because they are too weak and/or lack corresponding broadband images.

\section{New H$\alpha$ Emission-Line Stars}

\begin{figure*}
\centering
\includegraphics[width=18cm]{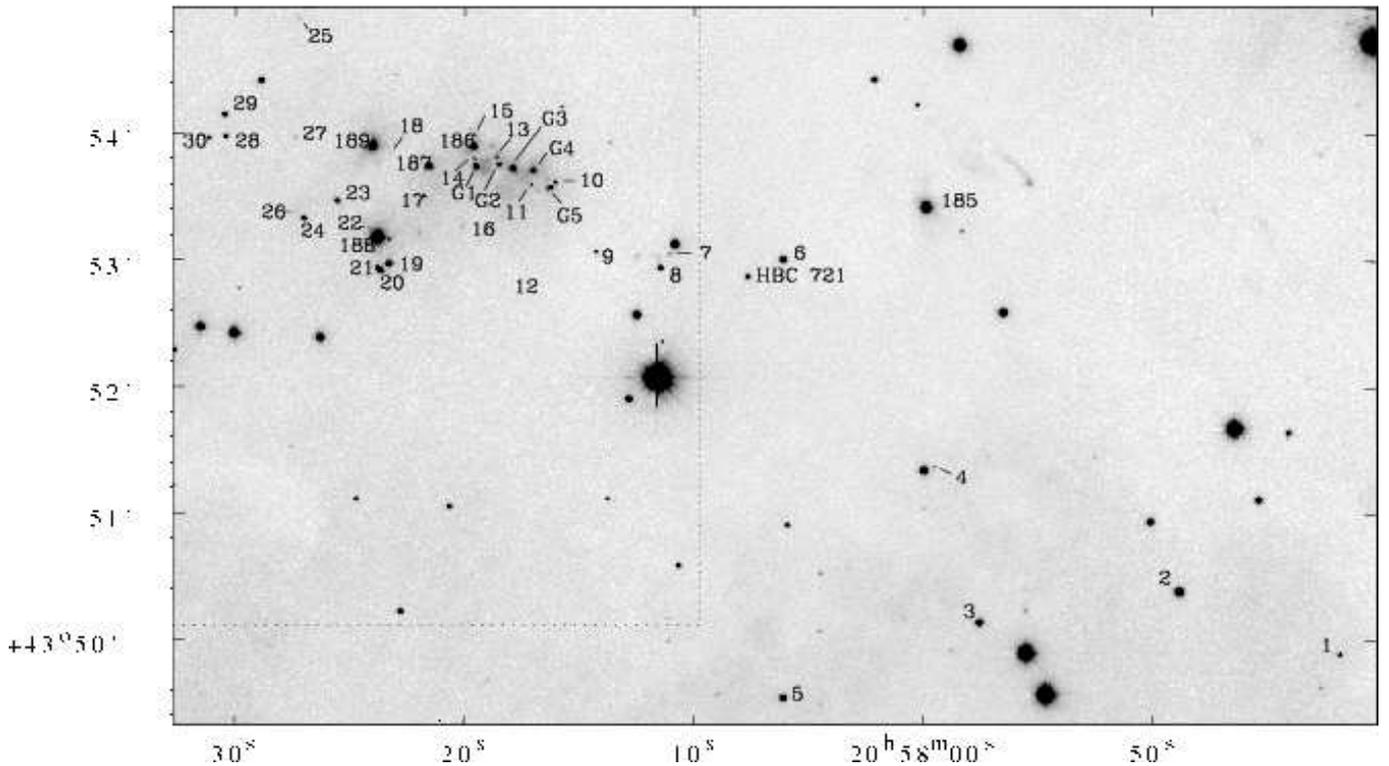}
\caption{$R$ band image showing the \halpha emission-line stars found.
The numbers between 1 and 30 are the MK\halpha identification from
the present work (Table~4). Stars with previous identifications
did not receive new numbers. 
Labels between 185 and 189 refer to LkH$\alpha$ numbers. 
The dotted lines mark the smaller field area of the grism images taken in 1998.}
\label{fig13}
\end{figure*}

In our grism images using the \halpha filter, the emission line is
approximately centered in the slitless spectra and it is rather easy to detect
the point-like emission. The images are deep and we can detect the
continuum of stars down to approximately $R$ = 21~mag.

A total of 41 \halpha emission-line stars were found in the region
surveyed (see Fig. 13 and Table~4), which corresponds to
approximately 50$\%$ of the stars visible in our grism images. They
are located mostly around the Lk\halpha 186 cluster, but there are
also \halpha emission-line stars to the west. $90\%$ of the stars
within 2\arcmin\ from Lk\halpha 189 show emission in H$\alpha$.

Some of the stars were known from Herbig (1958), Cohen \& Kuhi (1979)
and Kohoutek \& Wehmeyer (1999), listed in Table~1.
The previously unknown emission-line stars are identified by us
with MK\halpha numbers (MK stands for Mauna Kea).

Among the stars from Kohoutek \& Wehmeyer (1999) corresponding to the
IRS stars from Cohen \& Kuhi (1979), IRS~3 (KW~53-11) is identified in
both catalogs as a star around magnitude 17.  We do not see emission
from this star, but detected \halpha emission in a fainter star, which
we called MK\halpha 4, with magnitude 22, just northeast from the
brighter star.  IRS~4 is outside of our grism field and IRS~6 was -- at
the time of our observations -- optically too faint to be detected. We
could only find a correspondence to IRS~5, the optically visible
HBC~721.  The other optically visible stars with an \halpha emission
line listed in Table~1 were all detected in our survey.

The photometric survey from Laugalys et al. (2006) covers the entire
region of the Gulf of Mexico with a limiting magnitude of 
$V$ = 17.5~mag. They suspect the presence of \halpha emission in 
nine stars in the region surveyed by us, including Lk\halpha 185, 188, 189,
MK\halpha 2, 3 and 5, in all of which we detect H$\alpha$ emission.
The other three stars were also checked on our grism images and no
emission lines were seen (stars II-109, 113 and 118 in their Table~2).
Their limiting magnitude prevented them from detecting other known
emitters in the area, such as Lk\halpha 186 and 187.  
In their recent spectroscopic follow-up, Corbally et al. (2009)
confirm the \halpha emission lines also seen by us in MK\halpha 2, 3 and 5
and determine their spectral types (M3.5e, M1e and K7e, respectively).
They put MK\halpha 2 at a distance of 137 pc from the Sun,
excluding it from the star formation complex.
MK\halpha 3 and 5 might have distances compatible with the complex.
They classify II-109 and II-113 as G-type stars with \halpha filled 
in with emission, which explains why we did not detect the emission 
in our grism images.

In first-epoch grism images obtained in 1998 by George Herbig and
kindly put at our disposal, almost all the \halpha emission-line stars
were also detected (25 out of the 33, in the common field of view).
The equivalent widths of the lines were measured at both epochs and in
almost all of the cases the H$\alpha$ emission line strength is
comparable (see Table~4).

In the recent Subaru images obtained in very good seeing it is
possible to resolve three of the stars in the cluster as doubles, with
angular separations of less than 1\arcsec: G1, G3 and MK\halpha 14.
Because of the limited resolution in the grism images, it is not
possible to confirm if both components in each system have emission,
but we see tentative evidence that this is true at least in the case
of MK\halpha 14.

\begin{table*}
\caption{\halpha emission-line stars in the Gulf of Mexico.}
\label{tab4}
\centering
\begin{tabular}{cccccccccccc}
\hline\hline
MK\halpha & Alt. Name$^a$ & $\alpha$(2000) & $\delta$(2000) & $V$ & $R$ & $I$ & $J^b$ & $H^b$ & $K_s^b$ & W$_{(2002)}^c$ & W$_{(1998)}^c$ \\
\hline
  1  &               & 20 57 41.8 & +43 49 54 & 20.85 & 19.37 &  --   & 13.97 & 13.16 & 12.86      &    9 &    \\
  2  & L III-1       & 20 57 48.9 & +43 50 24 & 17.17 & 16.18 &  --   & 12.56 & 11.96 & 11.73      &   12 &    \\
  3  & L II-114      & 20 57 57.5 & +43 50 10 & 17.82 & 16.85 & 16.28 & 13.84 & 13.14 & 12.91      &    7 &    \\
  4  &               & 20 57 59.5 & +43 51 23 & 22.17 & 20.75 & 18.94 & 15.70 & 14.78 & 14.34      &    8 &    \\
     & Lk\halpha 185 & 20 57 59.9 & +43 53 26 & 16.22 & 15.08 & 14.60 & 11.85 & 10.87 & 10.25 $^+$ &  219 &    \\
  5  & L II-122      & 20 58 06.1 & +43 49 34 & 18.32 & 17.28 & 16.39 & 13.89 & 13.12 & 12.78      &   46 &    \\
  6  &               & 20 58 06.2 & +43 53 01 & 17.93 & 16.74 & 15.95 & 13.41 & 12.58 & 12.21      &   29 &    \\
     & HBC 721       & 20 58 07.6 & +43 52 53 & 20.55 & 18.90 & 17.52 & 14.10 & 12.79 & 11.98 $^+$ &   50 &    \\
  7  &               & 20 58 11.1 & +43 53 04 & 22.86 & 20.97 & 18.66 & 14.79 & 13.86 & 13.38      &   13 & -- \\
  8  &               & 20 58 11.4 & +43 52 58 & 16.63 & 16.43 & 16.14 & 13.75 & 12.21 & 11.21 $^+$ &  104 & 97 \\
  9  &               & 20 58 14.3 & +43 53 05 & 20.89 & 19.44 & 17.53 & 14.17 & 13.39 & 12.94      &   35 & 45 \\
 10  &               & 20 58 16.1 & +43 53 37 & 21.22 & 19.38 & 17.43 & 13.46 & 11.33L& 10.58L     &   25 & 26 \\
     & G5            & 20 58 16.3 & +43 53 35 & 20.24 & 18.53 & 16.94 & 13.03 & 11.21L& 10.42L     &   17 & 22 \\
     & G4            & 20 58 17.0 & +43 53 43 & 19.09 & 17.67 & 16.55 & 13.25 & 12.21 & 11.46 $^+$ &   69 & 68 \\
 11  &               & 20 58 17.1 & +43 53 36 & 21.77 & 20.15 & 18.00 & 14.09L& 13.00L& 12.96      &   13 & 22 \\
     & G3 *          & 20 58 17.9 & +43 53 44 & 18.58 & 16.84 & 15.48 & 11.77 & 10.42 &  9.70      &   29 & 11 \\
 12  &               & 20 58 18.4 & +43 52 48 & 23.47 & 22.02 & 19.53 & 15.63 & 14.91 & 14.32 $^+$ &    4 & -- \\
     & G2            & 20 58 18.5 & +43 53 47 & 20.53 & 18.75 & 17.08 & 13.21L& 12.18 & 11.39      &   26 & 30 \\
 13  &               & 20 58 18.6 & +43 53 49 & 21.83 & 19.90 & 18.30 & 13.39L& 12.76 & 11.96      &    5 &  9 \\
     & G1 *          & 20 58 19.5 & +43 53 45 & 19.34 & 17.65 & 16.18 & 12.62 & 11.35 & 10.84      &    7 & -- \\
 14  &    *          & 20 58 19.6 & +43 53 48 & 22.71 & 20.58 & 18.90 &  --   &  --   &  --        &    5 & 15 \\
 15  &               & 20 58 19.6 & +43 54 00 & 21.70 & 20.13 & 18.08 & 13.79L& 13.69 & 13.29      &    4 & -- \\
     & Lk\halpha 186 & 20 58 19.6 & +43 53 55 & 18.12 & 16.68 & 15.85 & 12.69 & 11.49 & 10.92      &   30 & 33 \\
 16  &               & 20 58 20.1 & +43 53 17 & 23.30 & 21.60 & 19.49 & 15.93 & 14.85 & 14.48      &    7 & -- \\
     & Lk\halpha 187 & 20 58 21.6 & +43 53 45 & 18.56 & 17.05 & 16.00 & 12.80 & 11.49 & 10.71 $^+$ &  162 & 40 \\
 17  &               & 20 58 21.7 & +43 53 31 & 22.40 & 20.62 & 18.17 & 13.77 & 12.36 & 11.80      &    9 & -- \\
 18  &               & 20 58 23.3 & +43 53 51 &  --   & 22.27 & 18.10 & 15.82 & 14.88 & 14.41      &    6 & nc \\
 19  &               & 20 58 23.3 & +43 52 59 & 17.81 & 16.62 & 15.65 & 12.77 & 11.81 & 11.34      &   40 & 25 \\
 20  &               & 20 58 23.7 & +43 52 56 & 18.30 & 16.98 & 15.87 & 12.87 & 12.07 & 11.74      &   24 & 12 \\
 21  &               & 20 58 23.9 & +43 52 57 & 18.62 & 17.29 & 16.06 &  --   &  --   &  --        &   10 &  4 \\
     & Lk\halpha 188 & 20 58 23.8 & +43 53 12 & 14.65 & 13.69 & 13.37 & 10.56 &  9.62 &  8.84 $^+$ &   98 & 28 \\
     & Lk\halpha 189 & 20 58 24.0 & +43 53 55 & 17.10 & 15.80 & 15.03 & 12.23 & 11.16 & 10.69      &   92 & 29 \\
 22  &               & 20 58 24.4 & +43 53 15 & 20.79 & 19.56 & 18.31 &  --   &  --   &  --        &   15 & -- \\
 23  &               & 20 58 25.6 & +43 53 29 & 19.58 & 18.11 & 16.81 & 13.60 & 12.40 & 11.73      &   46 & 37 \\
 24  &               & 20 58 27.1 & +43 53 20 & 19.54 & 18.20 & 17.26 & 14.02 & 12.92 & 12.25 $^+$ &  144 & 45 \\
 25  &               & 20 58 27.3 & +43 54 55 & 23.65 & 21.81 & 19.98 & 15.87 & 15.03 & 14.40 $^+$ &    6 & nc \\
 26  &               & 20 58 27.4 & +43 53 24 & 23.60 & 21.78 & 19.86 & 16.08 & 15.03 & 14.34 $^+$ &    5 & -- \\
 27  &               & 20 58 27.5 & +43 53 58 & 22.80 & 21.51 & 19.41 & 15.28 & 14.32 & 13.79      &    3 & 15 \\
 28  &               & 20 58 30.4 & +43 53 59 & 19.87 & 18.61 & 17.41 & 14.82 & 13.67 & 12.89 $^+$ &   70 & 42 \\
 29  &               & 20 58 30.5 & +43 54 10 & 20.68 & 19.20 & 17.63 & 13.79 & 12.84 & 12.41      &   69 & 17 \\
 30  &               & 20 58 31.2 & +43 53 58 & 19.10 & 17.78 & 16.63 & 15.33 & 14.04 & 13.29      &    4 &  5 \\

\hline
\end{tabular}
\begin{list}{}{}
\item[$^a$] -- Lk\halpha numbers from Herbig (1958); G numbers from Cohen \& Kuhi (1979); HBC number from Herbig \& Bell (1988); L numbers from Laugalys et al. (2006), with possible \halpha emission.
\item[$^b$] -- $JHK_s$ magnitudes extracted from the 2MASS All-Sky Catalog. All magnitudes marked L are upper limits.
\item[$^c$] -- H$\alpha$ equivalent widths measured in 2002 and 1998 images. 
{\em nc} means \halpha emission with no or very faint continuum.
\item[$^*$] -- Visual binaries.
\item[$^+$] -- Stars with infrared excess in the $JKH$ color-color diagram (Fig. 15).
\end{list}
\end{table*}

Table~4 lists the 41 \halpha emission-line stars identified.
The first column gives an MK\halpha (Mauna Kea) identification number 
to all the stars previously unknown as \halpha emitters. Column two gives
the identification for the previously known emission-line stars, as
well as other possible designations. Columns three and four give 
coordinates of the stars. The following three columns provide the 
optical $VRI$ magnitudes obtained in our observations,
while the next three columns list the near-infrared $JHK_s$ magnitudes
obtained from the 2MASS All Sky Survey Catalog. The \halpha emission
line equivalent widths measured in 1998 and 2002 are provided in the 
last two columns of the table. The optical binaries are also marked in
the table, as well as the stars showing near-infrared excess 
(see next section).

From the previously known spectral types of nine of the stars
we have determined expected main-sequence $(J-H)_0$ colors
from Bessell \& Brett (1988) and estimated extinction
values ($A_V$) for those stars. The same information
was also obtained by de-reddening the stars down to a location on
the main sequence in the $(J-H)\times(H-K_s)$ diagram shown in the next section.
The method is described in Herbig \& Dahm (2006). 
The values obtained using both methods agree within 0.6 mag.
The method of de-reddening was applied to all the stars
that have $JHK$ colors.

The extinction-corrected $V_0$ magnitudes and $(V-I)_0$ colors enabled
us to place the stars in a color-magnitude diagram, from which we can
get a rough estimate of their masses and ages. That diagram is shown in Fig.~14.
The evolutionary tracks and isochrones from D'Antona \& Mazzitelli (1997) 
were translated into the observational plane using the relationships of 
Hillenbrand (1997).
Most stars seem to have masses between 0.2 and 1~M$_{\sun}$, with
Lk\halpha 188 showing highest mass and MK\halpha 12 the lowest.
All the optically visible \halpha emission stars are older than 5 Myr. 
Note that the errors in the $A_V$ values can lead to errors of about 30\% 
in the estimate of their masses.

\begin{figure}
\centering
\includegraphics[width=8.8cm]{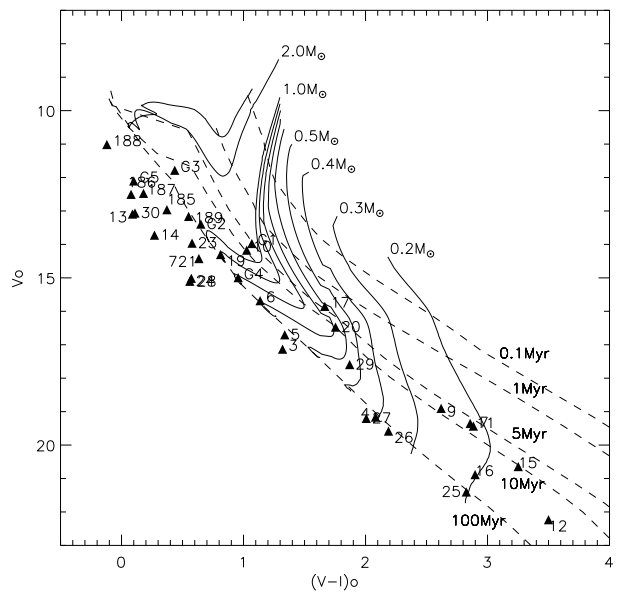}
\caption{Color-magnitude diagram for the \halpha emission-line stars (triangles).
An extinction-correction was applied to obtain $V_0$ and $(V-I)_0$ values.
The solid lines are evolutionary tracks from D'Antona \& Mazzitelli (1997),
for masses from 0.2 to 2.0 M$_{\sun}$ and the dashed lines are their
0.1, 1, 5, 10 and 100 Myr isochrones. The stars are labeled as in 
Fig.~13.}
\label{fig14}
\end{figure}

\section{Embedded Sources}

We use the $JHK_s$ 2MASS magnitudes of the stars in the field surveyed
to make a color-color diagram (Fig.~15) to identify stars with
infrared excess, indicating the presence of a disk.  The figure shows
the location of main sequence and giant stars from Bessell \& Brett
(1988) in solid and short-dashed lines respectively. 
The Classical T-Tauri Stars (CTTS) location from Meyer et al. (1997) 
is also indicated as a long-dashed line. A correction to the
2MASS photometric system was performed following the prescription of
Carpenter (2001).
The three parallel dotted lines show the direction of the interstellar
reddening vectors determined for the L935 region by Strai\v{z}ys et
al. (2008).  All the stars falling to the right of the middle vector
have clear infrared excess, they constitute about $25\%$ of the stars
in the figure.

\begin{figure}
\resizebox{\hsize}{!}{\includegraphics{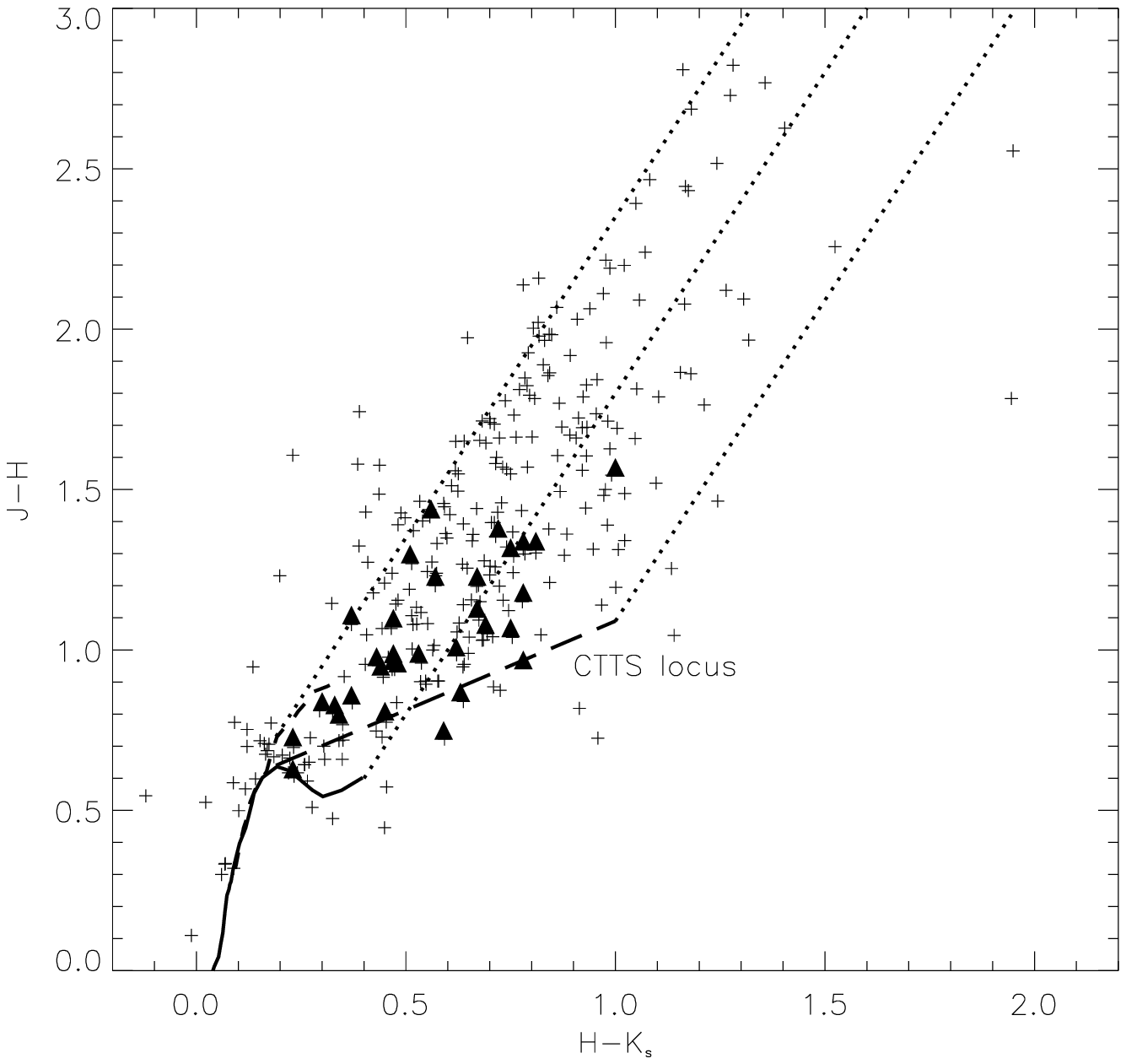}}
\caption{$JHK_s$ diagram of the stars detected in all the 3 bands with 2MASS
in the region surveyed.
Solid triangles are the \halpha emission-line stars in Table~4. The solid and 
dashed lines are, respectively, the location of main sequence 
and giant stars from Bessell \& Brett (1988) corrected to the 2MASS 
photometric system (Carpenter 2001). The long-dashed line represents the
CTTS location from Meyer et al. (1997). The dotted lines show the direction 
of the interstellar reddening vectors from Strai\v{z}ys et al. (2008).}
\label{fig15}
\end{figure}

The \halpha emission-line stars are marked as triangles in the $JHK_s$
diagram and 11 of them show some infrared excess. 
Those stars are marked in Table~4.
So, 70\% of the \halpha emission-line stars show colors typical 
of more evolved young 
stars with limited circumstellar material. 
Lk\halpha 188 is the emission-line star with larger infrared excess,
but apparently only little extinction.

8 out of the 10 stars with \halpha emission line equivalent width 
larger than 50~\AA\ have infrared excess. The other 3 stars with
infrared excess have a very small \halpha emission line equivalent 
width ($\le$ 6~\AA).

The $JHKL$ IRTF images cover only the central cluster area,
so we extracted near-infrared photometry only for those stars 
that fall inside the IRTF 75\arcsec\ field of view.
The magnitudes obtained are listed in Table~5, as well as the
coordinates, a near-infrared identification number (NIR Id.) 
and its correspondence to a MK\halpha number when one exists.
Figure~16 shows all the four IRTF images, with the identification 
numbers for the near-infrared stars detected, as in Table~5,
and also the corresponding {\it Spitzer} images at 8~$\mu$m and 24~$\mu$m.

A color-color diagram was plotted in a similar way to the one built
with 2MASS magnitudes (Fig.~17). This time the locations of main
sequence and giant stars were not corrected for the 2MASS photometric
system, as we use the standard IRTF $JHK$ filters. Among the 23 stars
detected in all the three bands, 9 have infrared excess ($39\%$). Our
observations are concentrated in a small area where it is likely that
almost all the stars are part of the young population. The relatively
high fraction of stars with little or no excess emission suggests that
the overall population is already several million years old.

\begin{figure}
\centering
\includegraphics[width=8.8cm]{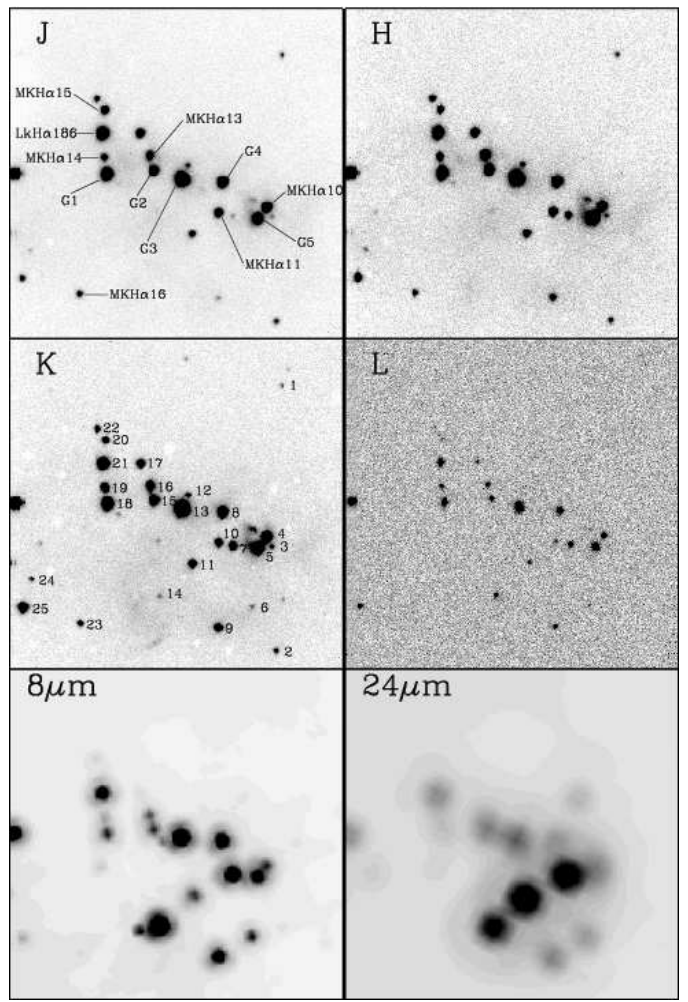}
\caption{$J$, $H$, $K$ and $L$ images, with the NIR identification
numbers for the Lk\halpha 186 cluster used in Table~5 shown in the $K$
image and previous designations shown in the $J$ image.  At the
bottom, the corresponding area as observed by {\it Spitzer} at
8~$\mu$m and 24~$\mu$m.  Each image is 75\arcsec\ on the side.}
\label{fig16}
\end{figure}

\begin{figure}
\resizebox{\hsize}{!}{\includegraphics{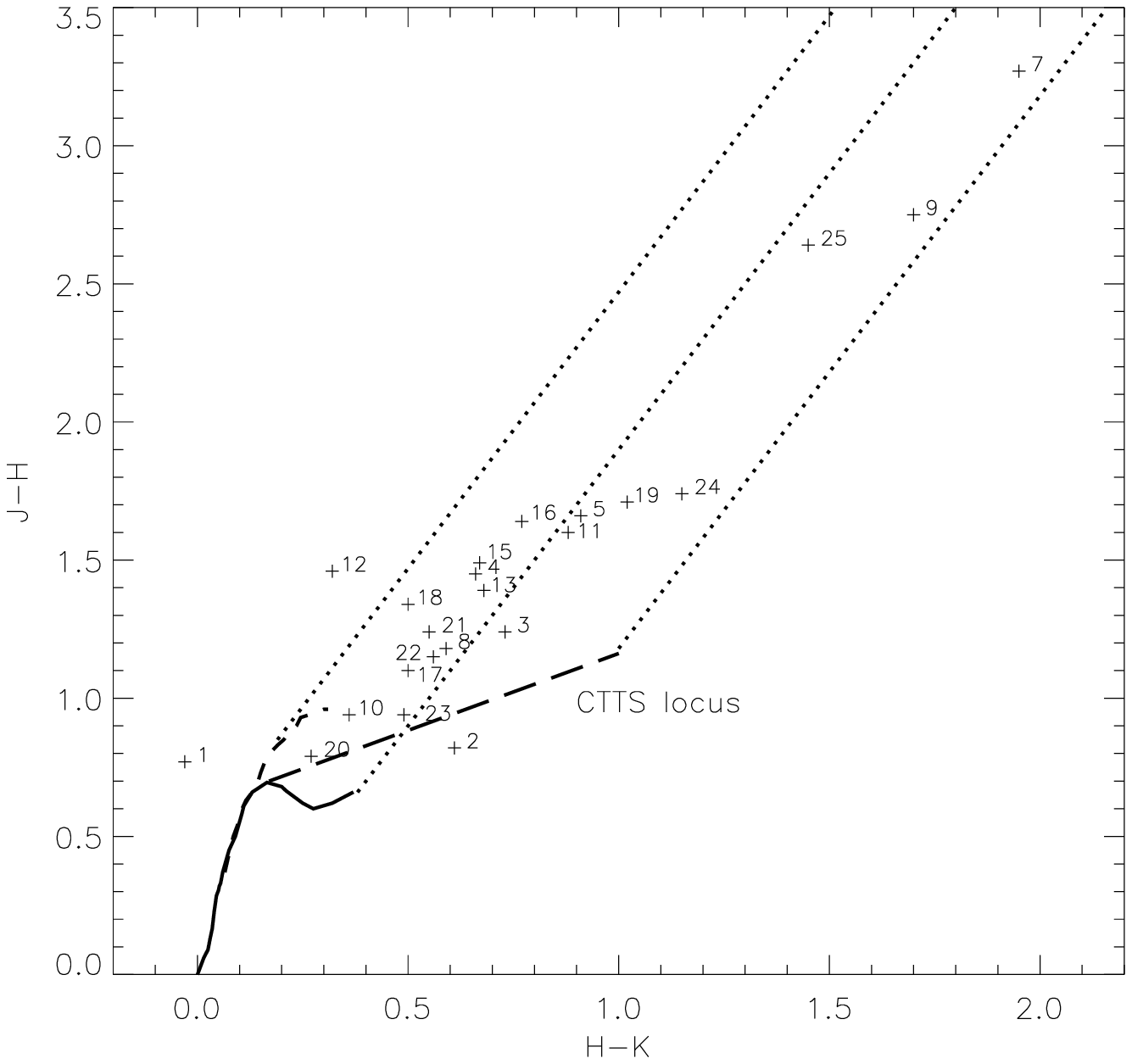}}
\caption{$JHK$ diagram of the Lk\halpha 186 cluster showing the stars 
detected in the IRTF $J$, $H$ and $K$ images. The solid and dashed 
lines are, respectively, the location of main sequence and giant stars 
from Bessell \& Brett (1988) and the long-dashed line represents the
CTTS location from Meyer et al. (1997). The dotted lines show the direction 
of the interstellar reddening vectors from Strai\v{z}ys et al. (2008).
The stars with infrared excess are marked in Table~5.}
\label{fig17}
\end{figure}

{\it Spitzer} magnitudes extracted for IRAC 3.6, 4.5, 5.8 and 8.0 $\mu$m
and MIPS 24 $\mu$m are listed in Table 6, for both the \halpha
emission-line stars and the NIR stars detected in the IRTF images.
A color-color diagram (Fig.~18) was built in order to classify 
the stars according to the regions they occupy in the diagram
(Allen et al. 2004). 
The majority of stars lie in the region of
Class II (Classical T-Tauri) stars.
Most of the Class I stars are possible sources of HH objects.
Stars that fall in the Class III (Weak-line T-Tauri) region
indeed show weak or no \halpha emission-line.

Near the optical cluster, the most embedded sources lie to its south,
as shown in the {\it Spitzer} 8~$\mu$m and 24~$\mu$m images 
(Fig.~16). At 24~$\mu$m NIR 7, 11 and 14 become significantly 
brighter than the others.
Star NIR 7 is located in the upper right corner of the $JHK$ 
color-color diagram, it is one of the stars with higher 
infrared excess and it is heavily extincted.  
It is barely seen at $J$ and becomes brighter as we move
to longer wavelengths. It is located near HH 654 and is likely its
source. In the {\it Spitzer} color-color diagram, it could be a Class II
star, given the amount of extinction it has. 
Star NIR 11 shows relatively large extinction, but no significant 
infrared excess in $JHK$.
Star NIR 14 is only seen at $K$ and $L$ and is close to a 
nebulosity seen at $K$ and an H$_2$ flow associated with 
HH~656. Its position indicates that it could be the source of this flow.
Both NIR 11 and NIR 14 are classified as Class I protostars, as well
as NIR 6, NIR 9, MK\halpha 8 and HBC 721. MK\halpha 8 is the
emission-line star with larger extinction in the $JHK_s$ diagram (Fig. 13).
In the $JHK$ diagram of Fig.~17, NIR 9 and NIR 25 also show a large
amount of extinction, both lie to the south of the optical cluster.
NIR 25 is located in the Class III region of the {\it Spitzer} 
color-color diagram.

\begin{table*}
\caption{$JHKL$ observations of the Lk\halpha 186 cluster.}
\label{tab5}
\centering
\begin{tabular}{cccccccc}
\hline\hline
NIR Id & \halpha Id & $\alpha$(2000) & $\delta$(2000) & $J$ & $H$ & $K$  & $L$  \\
\hline
 1 &               & 20 58 15.7 & +43 54 12 & 16.39 & 15.62 & 15.65      & -     \\
 2 &               & 20 58 15.8 & +43 53 12 & 16.15 & 15.33 & 14.72 $^+$ & -     \\
 3 &               & 20 58 16.0 & +43 53 35 & 16.07 & 14.83 & 14.10 $^+$ & -     \\
 4 & MK\halpha 10  & 20 58 16.1 & +43 53 37 & 13.49 & 12.04 & 11.38      & 8.21  \\
 5 & G5            & 20 58 16.3 & +43 53 35 & 13.36 & 11.70 & 10.79 $^+$ & 7.30  \\
 6 &               & 20 58 16.4 & +43 53 22 &  -    &  -    & 15.70      & -     \\
 7 &               & 20 58 16.8 & +43 53 36 & 17.49 & 14.22 & 12.27 $^+$ & 8.44  \\
 8 & G4            & 20 58 17.0 & +43 53 43 & 13.18 & 12.00 & 11.41      & 7.69  \\
 9 &               & 20 58 17.1 & +43 53 17 & 17.17 & 14.42 & 12.72 $^+$ & 9.06  \\
10 & MK\halpha 11  & 20 58 17.1 & +43 53 36 & 14.24 & 13.30 & 12.94      & -     \\
11 &               & 20 58 17.7 & +43 53 31 & 15.37 & 13.77 & 12.89 $^+$ & 9.61  \\
12 &               & 20 58 17.8 & +43 53 47 & 15.85 & 14.39 & 14.07      & -     \\
13 & G3            & 20 58 17.9 & +43 53 44 & 11.72 & 10.33 &  9.65      & 6.24  \\
14 &               & 20 58 18.4 & +43 53 24 &  -    &  -    & 15.68      & 8.89  \\
15 & G2            & 20 58 18.5 & +43 53 47 & 13.62 & 12.13 & 11.46      & 8.20  \\
16 & MK\halpha 13  & 20 58 18.6 & +43 53 49 & 14.38 & 12.74 & 11.97      & 8.95  \\
17 &               & 20 58 18.8 & +43 53 54 & 14.29 & 13.19 & 12.69      & 9.75  \\
18 & G1            & 20 58 19.5 & +43 53 45 & 12.66 & 11.32 & 10.82      & 7.86  \\
19 & MK\halpha 14  & 20 58 19.6 & +43 53 48 & 15.27 & 13.56 & 12.54 $^+$ & 9.16  \\
20 & MK\halpha 15  & 20 58 19.6 & +43 54 00 & 14.55 & 13.76 & 13.49      & -     \\
21 & Lk\halpha 186 & 20 58 19.6 & +43 53 55 & 12.63 & 11.39 & 10.84      & 7.52  \\
22 &               & 20 58 19.8 & +43 54 02 & 15.34 & 14.19 & 13.63      & -     \\
23 & MK\halpha 16  & 20 58 20.1 & +43 53 17 & 15.68 & 14.74 & 14.25      & -     \\
24 &               & 20 58 21.2 & +43 53 27 & 18.09 & 16.35 & 15.20 $^+$ & -     \\
25 &               & 20 58 21.3 & +43 53 21 & 15.69 & 13.05 & 11.60 $^+$ & 8.50  \\

\hline
\end{tabular}
\begin{list}{}{}
\item[$^+$] -- Stars with infrared excess according to the $JHK$ color-color diagram (Fig.~17).
\end{list}
\end{table*}

\section{Conclusions}

\begin{table}
\caption{IRAC and MIPS magnitudes for the \halpha emission-line stars and
the embedded NIR stars.}
\label{tab6}
\centering
\begin{tabular}{lcccccccc}
\hline\hline
 Id & [3.6] & [4.5] & [5.8] & [8.0] & [24] & Class.$^a$ \\
\hline
MK\halpha 1   & 12.40 & 12.16 & 11.78 & 11.17 &  7.96 & II  \\
MK\halpha 2   & 11.47 & 11.41 & 11.33 & 11.30 &   -   & III \\
MK\halpha 3   & 12.75 & 12.81 & 12.68 & 12.50 &   -   & III \\
MK\halpha 4   & 14.05 & 13.93 & 13.88 &   -   &  9.55 & III ? \\
Lk\halpha 185 &  9.36 &  8.80 &  8.35 &  7.52 &  3.84 & II  \\
MK\halpha 5   & 12.61 & 12.50 & 12.41 & 11.90 &  7.35 & II  \\
MK\halpha 6   & 11.48 & 11.23 & 10.75 &  9.90 &  6.93 & II  \\
HBC 721       & 10.38 &  9.49 &  8.75 &  7.98 &  5.20 & I   \\
MK\halpha 7   & 12.39 & 11.88 &   -   &     - &   -   & II ? \\
MK\halpha 8   &  9.31 &  8.51 &  7.62 &  6.69 &  3.18 & I \\
MK\halpha 9   & 12.12 & 11.69 & 11.32 & 10.85 &   -   & II  \\
MK\halpha 10  & 10.01 &  9.50 &  9.45 &  8.71 &   -   & II  \\
G5            &  9.11 &  8.65 &  8.64 &  7.99 &   -   & II  \\
G4            &  9.91 &  9.22 &  8.52 &  7.79 &   -   & I/II \\
MK\halpha 11  & 11.92 & 11.29 &   -   &     - &   -   & II ? \\
G3            &  8.83 &  8.36 &  7.86 &  7.23 &   -   & II  \\
MK\halpha 12  & 13.80 & 13.53 & 13.41 & 13.20 &   -   & III \\
G2            & 10.67 & 10.34 & 10.05 &  9.22 &   -   & II  \\
MK\halpha 13  & 11.00 & 10.50 & 10.07 &  9.21 &   -   & II  \\
G1            & 10.20 &  9.91 &  9.53 &  8.94 &   -   & II  \\
MK\halpha 14  & 11.43 & 10.94 & 10.61 &  8.93 &   -   & II  \\
MK\halpha 15  & 12.50 &  -    &   -   &     - &   -   &  -  \\
Lk\halpha 186 &  9.98 &  9.44 &  8.96 &  8.11 &  4.87 & II  \\
MK\halpha 16  & 13.40 & 12.93 & 12.48 & 11.95 &   -   & II  \\
Lk\halpha 187 &  9.41 &  8.84 &  8.31 &  7.59 &  5.03 & II  \\
MK\halpha 17  & 11.43 & 11.14 & 10.80 & 10.31 &   -   & II  \\
MK\halpha 18  & 13.24 & 12.98 & 12.51 & 11.73 &   -   & II  \\
MK\halpha 19  & 10.62 & 10.10 & 10.07 &  9.74 &   -   & II  \\
MK\halpha 20  & 11.19 & 10.87 & 10.43 &  9.51 &   -   & II  \\
MK\halpha 21  & 11.49 &   -   &  -    &  -    &   -   &  -  \\
Lk\halpha 188 &  7.88 &  7.43 &  7.05 &  6.27 &  3.60 & II  \\
Lk\halpha 189 &  9.98 &  9.56 &  9.31 &  8.75 &  6.18 & II  \\
MK\halpha 22  &   -   &   -   &  -    &  -    &   -   &  -  \\
MK\halpha 23  & 10.66 & 10.19 &  9.74 &  9.02 &  6.53 & II  \\
MK\halpha 24  & 11.21 & 10.79 & 10.55 &  9.77 &  7.27 & II  \\
MK\halpha 25  & 13.62 & 13.19 & 12.73 & 12.00 &  8.35 & II  \\
MK\halpha 26  & 13.39 & 12.95 & 12.44 & 11.39 &   -   & II  \\
MK\halpha 27  & 13.06 & 12.62 & 12.05 & 11.01 &  8.73 & II  \\
MK\halpha 28  & 11.71 & 11.10 & 10.54 &  9.93 &  7.34 & II  \\
MK\halpha 29  & 11.80 & 11.41 & 10.94 &  9.89 &  7.33 & II  \\
MK\halpha 30  & 12.05 & 11.74 & 11.49 & 11.00 &  7.62 & II  \\
NIR 1         & 15.27 & 14.97 & 14.48 & -     & -     & II ? \\
NIR 2         & 13.85 & 13.41 & 13.25 & 12.88 & -     & II  \\
NIR 3         &  -    &  -    &  -    &  -    & -     &  -  \\
NIR 6         & 12.02 & 10.56 &  9.70 &  8.88 & 4.61  & I \\
NIR 7         &  9.96 &  9.30 &  8.62 &  7.85 & 3.40  & I/II \\
NIR 9         & 10.79 &  9.58 &  8.88 &  8.02 & 4.51  & I \\
NIR 11        & 11.71 & 10.58 &  9.67 &  8.82 & 3.27  & I  \\
NIR 12        &  -    &  -    &  -    &  -    & -     &  -  \\
NIR 14        & 10.85 &  9.01 &  7.79 &  6.89 & 3.39  & I \\
NIR 17        & 11.37 &  -    & -     &  -    & -     &  - \\
NIR 22        & 12.55 & 12.28 &  9.74 &  8.96 & 4.87  & II  \\
NIR 24        & 11.89 & 11.39 & 10.58 & 10.33 & -     & II  \\
NIR 25        & 10.76 & 10.55 & 10.36 & 10.35 & -     & III \\

\hline
\end{tabular}
\begin{list}{}{}
\item[$^a$] -- Classification according to color-color diagram on 
Fig.~18. 
\item[?] -- Indicates the classification is based only in the
[3.6]-[4.5] color.
\end{list}
\end{table}

\begin{figure}
\centering
\includegraphics[width=9cm]{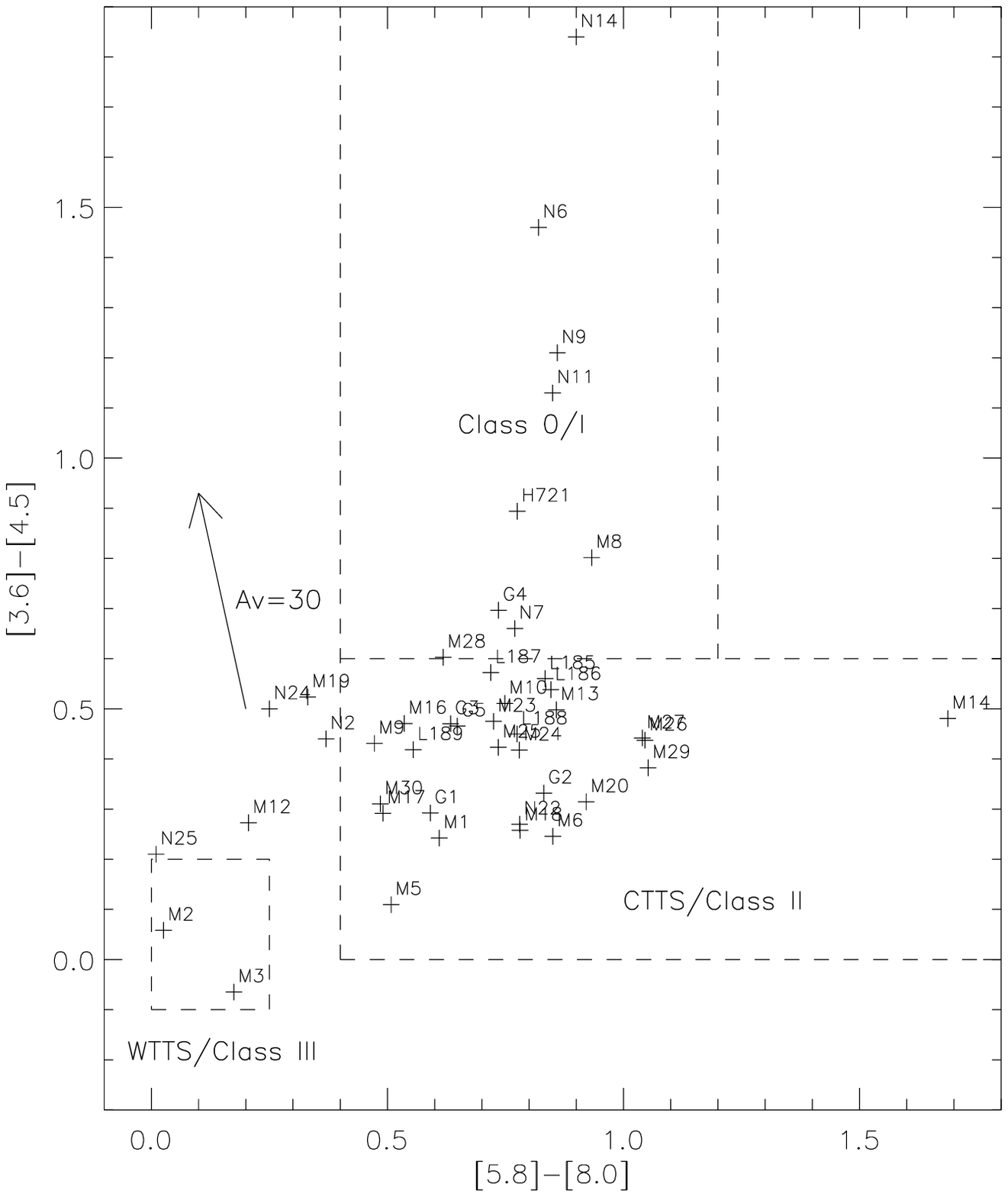}
\caption{An SPITZER/IRAC color-color diagram showing the positions of
the \halpha emission-line stars as well as the NIR stars. M stands
for MK\halpha emission-line stars, N stands for NIR stars, H stand for HBC
and G is the Cohen \& Kuhi (1979) designation for emitters in the 
Lk\halpha 188 cluster.}
\label{fig18}
\end{figure}

We have surveyed the region of the Gulf of Mexico around the little
optical cluster that contains Lk\halpha 185 to 189 using \halpha and \sii
filters and we found 35 new HH objects. The average
projected extent of the larger flows is approximately 1\arcmin, which,
at a distance of 550 pc, corresponds to $\sim$0.15~pc. 

Without proper motions it is rather difficult to identify the
source of each of the new flows. Images taken over several years may reveal
the overall motion of the HH objects, aiding in the identification of
their sources.  Our images span only from 2002 to 2006 and reveal no
sign of proper motion.
Based on our best resolution of 0.20~\arcsec/pix, we estimate that the
tangential velocities cannot be larger than 120 km/s, giving the 
assumed distance. This is similar to the typical velocities seen for HH flows.

A search for \halpha emission-line stars resulted in the detection of
many more young stars than previously known; 41 in the 14\arcmin
$\times$ 7\arcmin\ region surveyed, of which 30 are new. Also,
near-infrared images confirm that there are many embedded young
sources and flows in the region.  Classification based on {\it
  Spitzer} IRAC colors is provided for most of the sources. Almost all
the \halpha emission-line stars are Class II stars. Class I protostars
are found mainly among the near-infrared sources and are located south
and west of the optical cluster. Our observations reveal that star
formation is much more active in this area than previously suspected.

The exciting source of this entire HII region 
(2MASS J205551.25+435224.6), as identified by Comer\'on \& Pasquali 
(2005), lies directly to the west of the region studied, 
less than 25\arcmin\ away in projection, and presumably
only slightly more distant than the L935 cloud. 
Among the O-type candidates from Strai\v{z}ys \& Laugalys (2008), 
2MASS J205552.70+435324.2 lies very close to the Comer\'on \& Pasquali
(2005) source, and 2MASS J205806.73+435514.1 lies only 2\arcmin\ 
northwest from the optical cluster of Lk\halpha 188. 
It seems likely that the low-mass young stars we see represent 
second-generation star formation in the remnant clouds surrounding 
the W80 HII region which have been compressed by the central O-stars.

The optical cluster of the low mass \halpha emission-line stars is
only a small portion of the total number of stars being formed there,
and is only one of eight clusterings in the region according to the
recent {\it Spitzer} study by Guieu et al. (2009). The presence of
numerous HH flows and many reddened sources indicates that this is a
large site of widespread star formation, still partially embedded in
the dark cloud. Further studies in the Gulf of Mexico are encouraged.

\newpage

\begin{acknowledgements}
We are very grateful to George Herbig for placing his 1998 grism
images and measurements at our disposal.  
TA is also thankful for many discussions and suggestions from 
Luiz Paulo R. Vaz.  
We thank the referee Fernando Comer\'on for suggestions 
that improved this paper.
TA acknowledges financial support from CNPq/ Brazil under 
processes 200430/2001-7 and 201958/ 2007-4. 
BR was partially supported by the National Aeronautics
and Space Administration through the NASA Astrobiology Institute under
Cooperative Agreement No. NNA04CC08A issued through the Office of
Space Science, and by the NSF through grants AST-0507784 and
AST-0407005.  This work made use of observations from the {\em Subaru
Telescope}, which is operated by the National Astronomical Observatory
of Japan, from the {\it Spitzer Space Telescope}, which is operated by
the Jet Propulsion Laboratory, California Institute of Technology
under a contract with NASA, from the {\it Infrared Telescope
Facility}, which is operated by the University of Hawaii under
Cooperative Agreement No. NCC 5-538 with the National Aeronautics and
Space Administration, Science Mission Directorate, Planetary Astronomy
Program, and from the {\it United Kingdom Infrared Telescope}, which
is operated by the Joint Astronomy Centre on behalf of the Science and
Technology Facilities Council of the U.K.  We also acknowledge the use
of data products from the {\it Two Micron All Sky Survey}, which is a
joint project of the University of Massachusetts and the Infrared
Processing and Analysis Center/California Institute of Technology,
funded by NASA and NSF,
and Digitized Sky Surveys (DSS) images, produced at the Space 
Telescope Science Institute under U.S. Government grant NAG W-2166. 
\end{acknowledgements}


\end{document}